# Agentic AI and Human-in-the-Loop Interventions: Field Experimental Evidence from Alibaba's Customer Service Operations

Yiwei Wang[a], Chuan Zhu[b], Tianjun Feng[c], Lauren Xiaoyuan Lu[d], Bingxin Jia[e]

[a]Zhejiang University, International Business School, Zhejiang, China, yiweiwang@intl.zju.edu.cn

[b]Fudan University, School of Management, Shanghai, China, zhuc21@m.fudan.edu.cn

[c]Fudan University, School [1]of Management, Shanghai, China, tfeng@fudan.edu.cn

[d]Dartmouth College, Tuck School of Business, Hanover, NH 03755, lauren.x.lu@tuck.dartmouth.edu

[e]Alibaba Group Inc., bingxin.jbx@taobao.com

May 2026

**Abstract**

Agentic AI systems that autonomously perform service tasks are entering customer service operations. However, limited evidence exists on how human interventions shape service outcomes when agentic AI failures create both cognitive and emotional consequences. We study this issue through a randomized field experiment on Alibaba's Taobao platform. Workers in the treatment condition supervised an agentic AI system that resolved AI-eligible chats while continuing to handle AI-ineligible chats, whereas control workers resolved all chats without agentic AI. The findings show that AI deployment reduces average chat duration and has limited effects on retrial rates, but substantially lowers ratings for AI-eligible chats. Moreover, human intervention effectiveness in AI-eligible chats depends on the nature of AI failure, post-escalation intervention effort, and intervention timing. Human intervention preserves service quality in algorithm-triggered technical escalations, i.e., unresolved customer issues beyond the AI's capability, but is less effective in algorithm-triggered emotional escalations, i.e., where customers express frustration or dissatisfaction. These differences are partly explained by variation in workers' post-escalation intervention effort across escalation types. In algorithm-triggered emotional escalations, workers showed lower engagement: they sent fewer messages, contributed a smaller share of total chat rounds, and showed less proactivity in information seeking and solution provision. We further find that early intervention is essential for sustaining high post-escalation intervention effort. Finally, we document a positive spillover effect on AI-ineligible chats, as treated workers adapted their multitasking workflow to devote greater attention to these chats. These findings offer implications for human-in-the-loop process design in human-AI collaboration systems.

[1] The first two authors contributed equally.

## 1 Introduction

Recent advances in generative AI have given rise to a new class of agentic AI systems that can not only generate content but also execute tasks traditionally performed by human workers. In a recent survey of senior executives at leading companies, 35% of respondents reported having adopted AI agents by 2023, and another 44% indicated plans to deploy them in the near term (Ransbotham et al. 2025). This development has attracted substantial attention in customer service. Unlike earlier automation technologies, which supported workers at discrete stages of the service process, agentic AI can engage in turn-by-turn communication to interpret customer inquiries, determine appropriate next actions, and carry out service tasks in real time. However, the technology remains immature in important respects. In particular, agentic AI still faces limitations in adaptability and contextual judgment, which prevent it from reliably resolving all customer service requests. *Human-in-the-loop intervention* therefore remains an important safeguard. In this study, we examine how such intervention shapes service outcomes.

Our study departs in important ways from the existing literature on human-in-the-loop systems, which has largely focused on engineering settings such as medical imaging and warehouse operations. In these contexts, AI failures are predominantly cognitive in nature, and human intervention typically involves correcting well-defined errors (Sheridan and Parasuraman 2005, McKinney et al. 2020, Sun et al. 2022). By contrast, customer service represents a fundamentally different operational environment. The task is not only cognitive but also requires workers to manage customer emotions and sustain interpersonal rapport throughout the interaction. As a result, service failures in this setting may induce both cognitive and emotional consequences.

This distinction raises three unresolved questions: (1) How does deploying agentic AI under human-in-the-loop intervention affect customer service performance? (2) When and how can human interventions be effective? (3) How should firms design the process of human-agentic AI systems? Addressing these questions is critical for understanding the performance implications of deploying agentic AI in customer service operations.

We examine these questions using a randomized field experiment conducted on Alibaba's e-commerce platform, Taobao.com, in August 2024, involving 647 customer service workers and 680,676 service chats. At the time of the experiment, Alibaba had deployed an agentic AI system capable of autonomously resolving a subset of standardized service chats that were classified as AI-eligible. Otherwise, chats were classified as AI-ineligible, which required full human resolution. Workers assigned to the treatment condition supervised AI-eligible chats and intervened when necessary, while continuing to resolve AI-ineligible chats themselves. Workers in the control condition fully handled all chats throughout the study period.

Using worker-level difference-in-differences analysis, we find that the deployment of agentic AI reduces the average duration of chats and has limited effects on retrial rates, but substantially decreases customer ratings for AI-eligible chats. Furthermore, the results demonstrate that the effectiveness of human intervention on AI-eligible chats depends on the nature of AI failure, the post-escalation human intervention effort, and the timing of intervention.

Human intervention preserves service quality in algorithm-triggered technical escalations, i.e., unresolved customer issues beyond the AI's capability, but is less effective in algorithm-triggered emotional escalations, i.e., cases where customers have expressed frustration or dissatisfaction. These differences are partially explained by variations in the post-escalation intervention effort of human workers across different escalation types. For emotional escalations triggered by algorithms, workers exhibited lower engagement—they sent fewer messages, contributed to a smaller share of total chat rounds, and demonstrated less proactivity in information seeking and solution provision.

The timing of human intervention plays a central role in explaining when human-in-the-loop systems succeed or fail. In technical escalations, human workers intervene while the customer issue remains primarily a capability mismatch, leaving sufficient room for recovery. In emotional escalations, by contrast, intervention often occurs only after customer frustration has accumulated, making the interaction harder to repair and reducing workers' subsequent engagement. Human-initiated escalations suggest that earlier intervention can partially prevent this deterioration: when supervisors take over before negative sentiment becomes entrenched, they are better able to sustain effort and limit the decline in service quality.

Finally, we document a positive spillover effect on AI-ineligible chats, even though these chats were not directly handled by the agentic AI system. The results suggest that by offloading part of the regular workload to AI, treated workers reallocated attention toward AI-ineligible chats that required full human judgment. Consistent with this interpretation, multitasking analyses show that treated workers switched away from these focal chats less frequently and spent less time away from them. Across AI-eligible and AI-ineligible chats, the deployment of agentic AI significantly improves service speed but does not produce a material change in service quality.

These results have important implications for how firms design the process for human-AI collaboration in hybrid human-AI systems. Assigning workers to specialized supervisory roles may improve service quality in AI-eligible chats, but it may also deplete emotional resources and contribute to skill erosion over time. An integrated role, in which workers both supervise AI-eligible chats and fully resolve AI-ineligible chats, generates positive workload spillovers. At the same time, it exposes firms to lower customer ratings in emotionally escalated chats when intervention occurs too late. The appropriate process design therefore depends on how firms balance these competing considerations, and neither a fully specialized structure nor a fully integrated structure dominates along all dimensions.

The remainder of the paper is organized as follows. Section 2 reviews the related literature. Section 3 describes the experimental setting and the data. Section 4 presents the empirical strategy. Section 5 reports the results. Section 6 discusses managerial implications and concludes the paper.

## 2 Literature Review

### 2.1 Agentic AI in Service Operations

Recent advances in AI have shifted its role in service operations from an assistive tool to an autonomous agent. Prior research on AI in operations has primarily focused on prediction, reasoning, and content-generation tools that support human workers, including generative AI (Noy and Zhang 2023, Chen and Chan 2024, Brynjolfsson et al. 2025, Ni et al. 2026, Bai et al. 2026). In these settings, AI augments human labor by generating suggestions, drafting content, or synthesizing information, while humans remain the primary decision makers and communicators.

Agentic AI differs in an important way: rather than merely supporting human work, it can autonomously execute multistep tasks and directly interact with customers in real time. The central question is therefore no longer whether AI assistance improves human performance, but whether service systems perform effectively when agentic AI handles frontline labor and humans act as AI supervisors. Recent modeling work suggests that such systems may also reshape organizational structure by altering task allocation between different workers (Xu et al. 2025). Empirical evidence on these issues, however, remains scarce, particularly in customer-facing settings. This study contributes to addressing this gap by providing field experimental evidence from a real service environment in which agentic AI interacts directly with customers under human supervision. Our findings have important implications for how firms redesign the organizational structure and human labor in deploying agentic AI systems.

### 2.2 Human-in-the-Loop Interventions in Human-AI Service Systems

Secondly, we contribute to the literature on human-in-the-loop systems, which emphasizes the role of human intervention in improving system reliability by correcting AI errors and preserving accountability (Sheridan and Parasuraman 2005, McKinney et al. 2020, Sun et al. 2022, Lebovitz et al. 2022, Benjaafar et al. 2025). Most prior work, however, focuses on engineering settings such as autopilot systems, robotics, and medical AI, where failures are primarily cognitive and interventions are designed to correct well-defined errors.

Customer service differs in an important respect because failures of AI can have both cognitive and emotional consequences. When agentic AI interacts directly with customers, failures may accumulate through miscommunication and rising customer frustration before human intervention occurs. Consequently, the effectiveness of human intervention may depend not only on whether AI fails, but also on the type of failure and the timing of intervention. Prior studies show that customer responses to AI are shaped by features such as anthropomorphism, identity disclosure, and prior exposure to AI (Bai et al. 2022, Cui et al. 2022, Xu et al. 2024, Zhang and Narayandas 2026), but they provide limited evidence on how the impact of human intervention varies with the nature of AI failure. This study extends the literature by showing that the value of human-in-the-loop intervention is heterogeneous: it is more effective in algorithm-triggered technical escalations and human-initiated escalations, but substantially less effective in algorithm-triggered emotional escalations where intervention occurs only after customer sentiment has already deteriorated.

### 2.3 Emotional Labor in Customer Service Operations

Finally, we contribute to the literature on emotional labor in customer service operations. Classic research in service operations emphasizes the tradeoff between service speed and service quality, showing that customers value both timely responses and effective problem resolution, and that service failures often generate retrials and additional operational burden (Hu et al. 2022, Long et al. 2024). Research on online service support further shows that multitasking can increase throughput but also lengthen response delays and reduce service quality (Goes et al. 2018). A central feature of customer service, however, is that performance depends not only on cognitive problem solving, but also on emotional labor. Workers must regulate their communication to maintain rapport and manage customer frustration. Prior work shows that negative customer emotion increases service load and that emotional deterioration can become self-reinforcing over the course of a chat (Altman et al. 2021).

This setting becomes especially important in hybrid human-AI service systems. When agentic AI serves as the initial point of contact, failures may leave behind not only unresolved issues but also chats with substantial emotional load before a human worker intervenes. As a result, the effectiveness of subsequent human intervention may depend not only on whether the underlying issue can still be resolved, but also on the customer's emotional state. Prior studies show that the performance of AI in customer service is highly context dependent. Voice-based AI can reduce complaints in some cases, but recognition failures increase demand for human service and worsen customer outcomes (Wang et al. 2023). Similarly, AI-assisted online chat can improve response speed and customer sentiment in some chats but performs poorly when issues require recovery from prior service failures (Zhang and Narayandas 2026). We extend this literature by showing that human intervention performs substantially worse in emotionally escalated chats, in part because human workers become less engaged following intervention. Once a chat carries substantial emotional load, the scope for recovery narrows, and workers exert less effort to move the chat toward resolution.

## 3 Experiment Setup

### 3.1 Company Background

This study was conducted in partnership with Alibaba, the world's largest e-commerce platform by gross merchandise value in 2023. We examine a randomized field experiment implemented on Taobao, the primary online marketplace of Alibaba and a platform comparable to Amazon Marketplace in scale and functionality. To strengthen customer loyalty and encourage repeat purchases, Taobao operates a large customer support system that addresses a broad range of service issues. By 2024, this system employed approximately 38,000 service workers and served more than 380 million daily active users, handling up to one million chats per day during major promotional events. The scale, operational complexity, and customer-facing nature of this setting make it well suited to the study of agentic AI in service operations.

### 3.2 Online Customer Service at Alibaba

In this setting, customer service workers operated as independent remote gig workers who resolved customer inquiries individually. All service interactions took place on the Taobao platform, where workers communicated with customers through mobile devices or desktop interfaces. A service chat is defined as the complete sequence of messages exchanged between a customer and a worker within a single interaction. Workers were compensated under a piece-rate scheme. Base pay was proportional to the total number of chats resolved during a shift and was multiplied by a quality-based bonus coefficient determined by service performance.

Before the experiment, customer inquiries were first triaged by a rule-based chatbot that classified issue types. If customer issues were not resolved by the chatbot, they would be subsequently routed to human workers, who resolved the issues through turn-by-turn messaging while following standardized operating procedures (SOPs). These SOPs took the form of decision trees that specified predefined responses and recommended actions for common service scenarios.

Upon completion of each chat, customers were invited to rate the service experience on a five-point scale, with higher scores indicating greater satisfaction.

### 3.3 Deploying Agentic AI with Human Supervision

In May 2024, Alibaba developed an agentic AI system to support online customer service. The system uses a large language model (LLM) as its core decision engine to coordinate multiple generative AI functionalities and autonomously manage service chats. The deployment also introduced a redesigned process for handling online service chats, as illustrated in Figure 1.

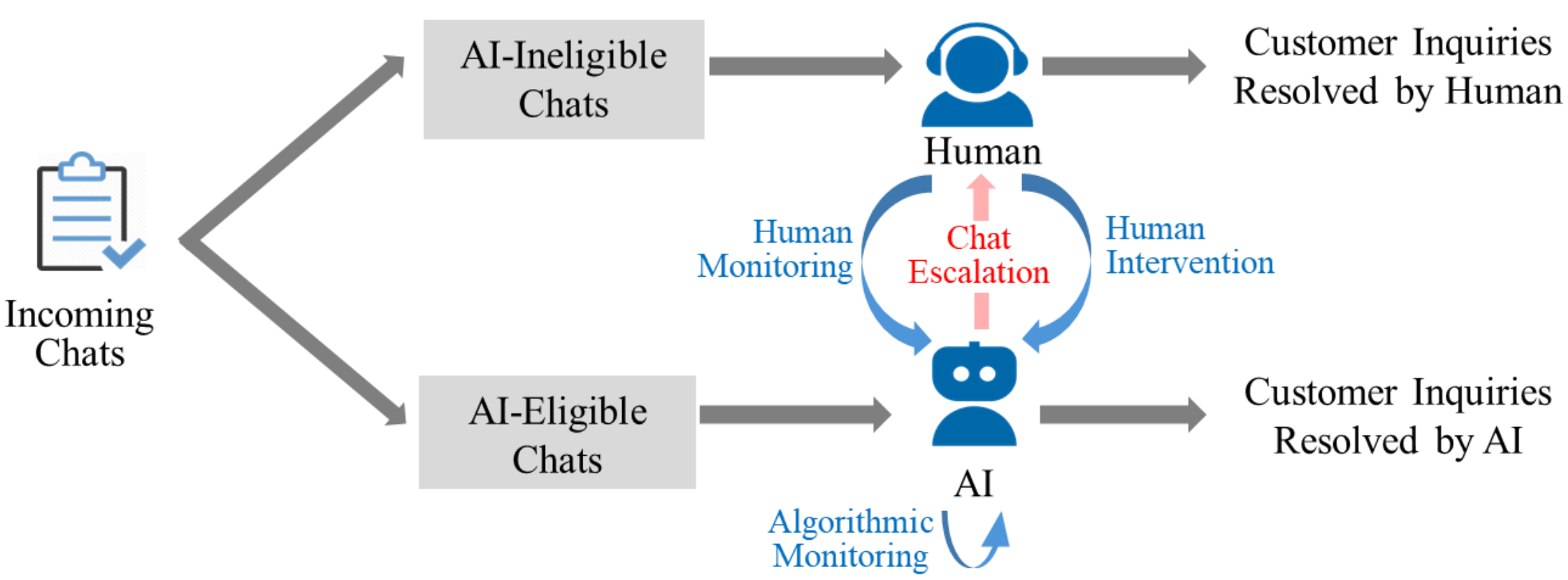

**Figure 1: Customer Service Process with the Agentic AI System**

Following the deployment of agentic AI, incoming chats were classified as either *AI-eligible* or *AI-ineligible*. AI-eligible chats involved standardized customer issues that could be resolved autonomously by the agentic AI system, whereas AI-ineligible chats involved non-standardized issues that required full human resolution.

At the time of the experiment, AI-eligible chats accounted for less than 10 percent of total chat volume. Even within this subset, the agentic AI system could not reliably complete all chats because of limitations in adaptability and interpersonal communication. To address these limitations, Alibaba embedded the agentic AI system within a human-in-the-loop structure. Each AI-eligible chat was assigned to a designated human worker who supervised the interaction and retained responsibility for service completion.

The supervisory role of human workers involved two core functions: *monitoring* and *intervention*. Supervisors could monitor AI-eligible chats through a dedicated interface. Because human monitoring capacity was limited, Alibaba also implemented algorithmic monitoring tools that continuously tracked the workflow of the AI system and flagged chats with an elevated risk of service failure. When such risk was detected by the algorithm, the chats were escalated to human supervisors. Alternatively, human supervisors may trigger escalations by themselves based on their own judgment. Once escalations are triggered, human supervisors then intervene in the workflow of the AI system, take over the chats, and complete the chats.

### 3.4 Experiment Design

Alibaba conducted a randomized field experiment from August 15 to August 31, 2024, to evaluate the impact of deploying the agentic AI system. The experiment involved 647 randomly selected customer service workers and generated 680,676 online service chats during the study period. Workers were randomly assigned to treatment based on their employee ID, resulting in 345 workers (53.3%) in the control group and 302 workers (46.7%) in the treatment group. Random assignment at the worker level ensures baseline comparability across groups, which we verify through balance checks as reported in Section 3.5.

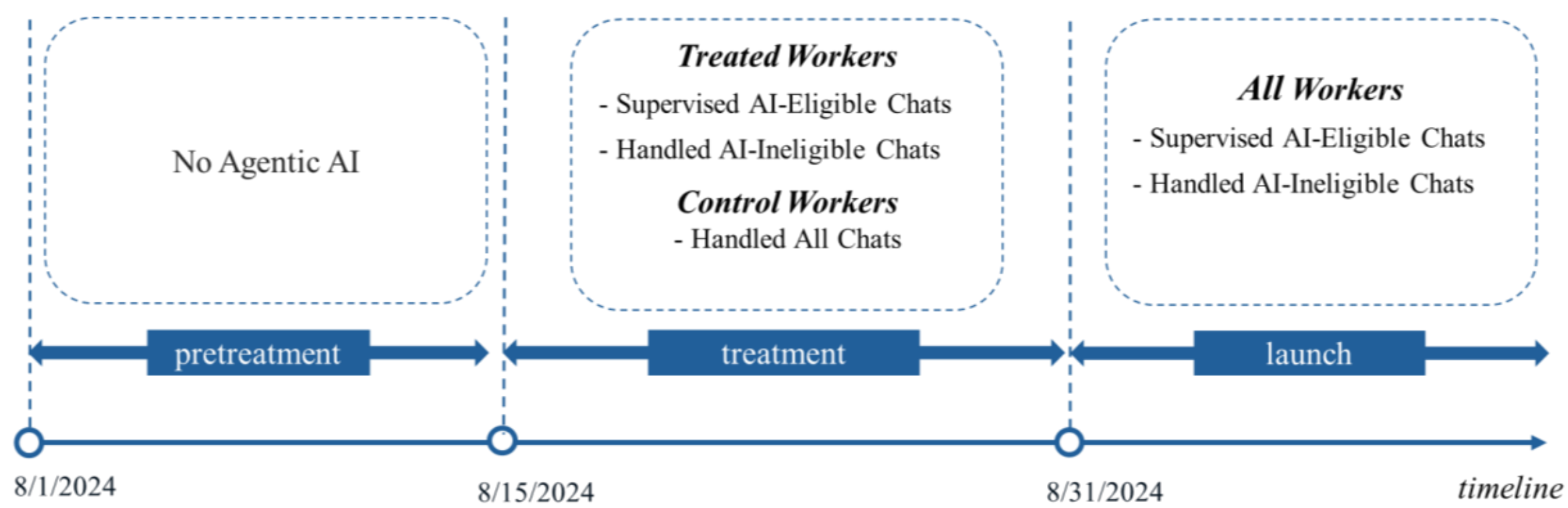

**Figure 2: Experiment Timeline**

Figure 2 presents the timeline of the experiment. The first 14 days, from August 1 to August 14, 2024, constitute the pretreatment period, during which the agentic AI system was not yet deployed and all service chats were handled by human workers. The following 17 days, from August 15 to August 31, 2024, constitute the treatment period, during which the agentic AI system was deployed exclusively for workers in the treatment group. During this period, treated workers supervised AI-eligible chats while continuing to handle AI-ineligible chats themselves. By contrast, control workers fully handled all chats. The experiment ended on August 31, 2024, after which the agentic AI system was rolled out to all workers beginning the next day.

Compensation under the redesigned process also varied by chat type. For AI-ineligible chats, workers continued to receive piece-rate compensation based on chat volume, multiplied by a quality-based bonus coefficient. For AI-eligible chats, workers received compensation based on chat volume regardless of whether the chat was handled by the agentic AI system under human supervision or by human workers alone. Regardless of who eventually resolved the customer issue, standardized operating procedures (SOPs) are used. AI-eligible and AI-ineligible chats also contributed equally to workload assessments.

### 3.5 Data Description

Our analysis integrates four primary data sources: (1) worker demographic records, (2) session logs, (3) chat transcripts, and (4) worker activity logs. Demographic records contain age, gender, and tenure. Session logs include worker and customer identifiers, session IDs, issue categories, and time stamps of messages, where a message is defined as a discrete text block submitted when the user presses the enter key. These logs also record customer satisfaction ratings and subsequent follow-up contacts. Chat transcripts provide the full textual content of each conversation, and activity logs capture time-stamped worker behaviors such as clicking, typing, or viewing conversation or order histories.

Merging these sources yields a chat-level dataset of 680,676 chats, of which 115,243 include customer ratings. Table 1 reports summary statistics for the key variables used in the analysis. From this integrated dataset, we construct several categories of service measures described below. Figure A2 in the Online Appendix illustrates selected key measures using a hypothetical chat.

***Demographics Measures.*** The average worker tenure is 12.5 months, and 19 percent of workers are male. As shown in Figure A1 in the Online Appendix, most workers joined the firm approximately eight to nine months prior to the experiment.

***Service Performance Measures.*** These measures capture the two primary operational objectives of customer support: service speed and service quality. Service speed is captured by *chat duration*, defined as the time elapsed between chat initiation and termination; the average chat lasts 462.36 seconds. Service quality is captured by both *customer rating* and *retrial*. Rating is measured on a five-point scale, with five being the highest rating. In our sample, 17% of chats received a rating with an average score of 3.57. Retrial is an indicator equal to one if the customer contacts the platform again regarding the same issue within seven days; 43% of chats result in a retrial. Rating response rates remain stable and comparable across treatment and control groups throughout the experiment (Figure A3, Online Appendix A).

***Chat Process Measures.*** We construct *total chat rounds*, *total message count*, *total word count*, and *response delays*, which are commonly used in text-based customer service to capture service effort and process dynamics (Altman et al. 2021, Zhang and Narayandas 2026). Specifically, total chat rounds represent the back-and-forth message exchanges between the service agent (human worker or AI agent) and the customer. On average, each chat involves 7.24 rounds, 13.82 messages, and 202.35 words. We also compute average response delay and total response delay, defined as the mean and cumulative time elapsed between a customer message and the service provider's subsequent reply. Across all chats, the average response delay is 14.76 seconds, and the cumulative delay is 107.19 seconds.

***Multitasking Measures.*** In online customer support, workers often serve multiple customers simultaneously (Long et al. 2024). To capture multitasking intensity, we construct three measures from overlapping chat timelines: *concurrency*, defined as the number of simultaneous chats at the start of the focal session; *total active time*, defined as the total time the worker is actively engaged with the focal chat interface; and *total away time*, defined as the total time the worker spends away from the focal chat interface. On average, workers handle 3.14 concurrent chats, with 248.07 seconds of active time per chat and 212.83 seconds of time away from the focal chat.

***Balance Check.*** In this setting, treatment is randomized at the worker level, whereas the primary operational data are observed at the chat level. To align the level of treatment assignment with the level of analysis, we aggregate chat-level measures to the worker-day level and construct panel datasets for subsequent difference-in-differences estimation to explore within-worker variation over time. To assess comparability, we examine 14 variables covering worker demographics, service performance, chat process characteristics, and multitasking behavior during the pretreatment period from August 1 to August 14. As reported in Table A1, nonparametric tests reveal no systematic differences between treated and control workers, supporting the validity of the randomization. Figure A3 indicates that customer rating response rates remain stable and comparable both before and after the treatment, and Figure A4 shows nearly identical discrete distributions of customer ratings across groups during the pretreatment period. Taken together, these results support the validity of the experimental design and indicate that posttreatment differences can be causally attributed to the deployment of the agentic AI system.

## 4 Empirical Strategy

This section defines the dependent and independent variables and presents the empirical framework used to estimate the treatment effects.

### 4.1 Variable Definitions

As described in Section 3.5, the underlying operational data are recorded at the chat level, whereas treatment is randomly assigned at the worker level. We aggregate chat-level measures to the worker-day level and construct three parallel worker-day panel datasets. The all-chats panel includes observations for all chats handled or supervised by a worker on a given day and serves as the main sample (680,676 chats).[2] The AI-eligible panel aggregates observations using only chats classified as AI-eligible (39,432 chats, 5.8 percent of the main sample), whereas the AI-ineligible panel aggregates observations using only chats classified as AI-ineligible (641,244 chats, 94.2 percent of the main sample). We analyze these three panels separately to estimate the total effects, direct effects, and spillover effects of deploying the agentic AI system.

Our primary dependent variables capture service performance, which we conceptualize along two dimensions: service speed and service quality. Service speed is measured by chat duration $(P_d)$, a standard operational measure of efficiency. Speed alone, however, does not fully capture service performance, because shorter duration may reduce labor input while also limiting opportunities to deliver high-quality service (Goes et al. 2018, Altman et al. 2021). We therefore examine service quality using both subjective and objective measures. Subjective quality is captured by customer rating $(Q_c)$, which reflects the customer's overall evaluation of the service experience. Objective quality is captured by retrial rate $(Q_r)$, an important operational measure within Alibaba, which has also been used in prior research as a proxy for ultimate issue resolution (Ni et al. 2026).

Our key independent variable is $Treat_{it}$, an indicator equal to one for treated workers during the treatment period and zero otherwise. In the empirical specifications, we also include worker fixed effects and calendar-day fixed effects to account for time-invariant worker heterogeneity and common temporal shocks, as described in Section 4.2.

Finally, we include several time-varying control variables that capture customer mix and workload conditions. These controls include lagged average customer tenure, lagged VIP customers served, lagged average concurrency, and lagged chat volume. We lag these variables by one day to mitigate simultaneity concerns, because customer composition and workload conditions may be jointly determined with service outcomes measured on the same day.

### 4.2 Empirical Specification

To estimate the average treatment effects of deploying the agentic AI system, we estimate the following difference-in-differences model:

$$Y_{it} = \beta_0 + \beta_1\, Treat_{it} + \theta_i + \mu_t + W_{i,t-1} + \epsilon_{it}, \tag{1}$$

Here, $Y_{it} \in \{\ln(P_d), Q_c, Q_r\}$ denotes the performance outcome for worker $i$ on day $t$ for a given chat category. We log-transform chat duration $P_d$ due to its relatively high variance. The indices

[2] In our sample, all chats are clearly labeled as AI-eligible or AI-ineligible for both treated and control workers.

$i$ and $t$ denote workers and calendar days, respectively. The terms $\theta_i$ and $\mu_t$ denote worker and calendar-day fixed effects, and $\epsilon_{it}$ is the idiosyncratic error term. $W_{i,t-1}$ captures lagged time-varying controls that account for shifts in customer composition and workload. These controls include lagged average concurrency and chat volume, as well as lagged average customer tenure and proportion of VIP customers served by worker $i$.

The variable $Treat_{it}$ equals 1 if worker $i$ belongs to the treatment group and day $t$ falls within the treatment period, from August 15 to August 31, 2024, and zero otherwise. The coefficient $\beta_1$ identifies the average treatment effect on the outcome variable. In all specifications, robust standard errors are clustered at the worker level.

## 5 Empirical Results

This section reports the empirical findings on the impact of deploying an agentic AI system with human-in-the-loop interventions. Section 5.1 examines the effects of agentic AI deployment on service performance. Section 5.2 analyzes the effectiveness of human-in-the-loop interventions in AI-eligible chats. Section 5.3 discusses managerial implications for AI supervision and process design. Section 5.4 presents a series of robustness checks.

### 5.1 Effects of Agentic AI Deployment on Service Performance

We begin by examining the total effects of agentic AI deployment on service performance across all chats. Table 2 shows that the deployment of agentic AI improves service speed at the aggregate level, with average chat duration declining by 3.2 percent ($p < 0.001$). By contrast, the effects on service quality are limited. The estimated effect on objective service quality, measured by retrial rates, is close to zero and statistically insignificant, and the estimated effect on subjective service quality, measured by customer ratings, is likewise statistically insignificant. Taken together, these results indicate that the deployment of agentic AI improves overall service speed without producing a significant change in service quality.

We next distinguish between AI-eligible and AI-ineligible chats. With respect to *service speed*, the deployment of agentic AI shortens average chat duration in both categories. As shown in Table 3 Column (1), average chat duration declines by 16.8 percent in AI-eligible chats ($p < 0.001$). Table 4 Column (1) shows that average chat duration also declines by 1.8 percent in AI-ineligible chats ($p < 0.05$). These findings indicate that the speed gains from agentic AI deployment are broad based, although they are much larger in AI-eligible chats.

With respect to *objective service quality*, agentic AI deployment has limited effects on retrial rates in both categories. Table 3 Column (2) shows that the estimated effect on retrial rates in AI-eligible chats is statistically insignificant, and Table 4 Column (2) shows the same null result for AI-ineligible chats. Taken together, these findings suggest that the deployment of agentic AI does not materially affect the likelihood of ultimate issue resolution.

The pattern differs sharply, however, for *subjective service quality*. For AI-eligible chats, Table 3 Column (3) shows that customer ratings decline by 0.412 points relative to those of control workers ($p < 0.001$). For AI-ineligible chats, by contrast, Table 4 Column (3) shows that treated workers receive customer ratings that are 0.091 points higher than those of control workers ($p < 0.01$).

Overall, these results show that the performance consequences of agentic AI deployment are fundamentally heterogeneous across chat types. Improvements in service speed appear in both AI-eligible and AI-ineligible chats, but customer perceptions of service quality depend critically on the type of chat. In AI-eligible chats, the deployment of agentic AI substantially accelerates service completion, yet these gains in speed do not translate into a better customer experience. In AI-ineligible chats, by contrast, agentic AI deployment generates positive spillovers: AI-ineligible chats handled fully by human workers become modestly faster and receive higher customer ratings.

### 5.2 AI-Eligible Chats: When Does Human-in-the-Loop Intervention Work?

#### 5.2.1 Subsample Analysis and Matching Procedures

Section 5.1 shows that agentic AI improves service speed but results in lower customer ratings in AI-eligible chats. In customer service, AI failures are not purely cognitive; they may also involve emotional deterioration during the chat. We therefore turn to the following question: *when and how can human-in-the-loop intervention preserve service quality for AI-eligible chats?* To answer this question, we examine the 11,069 AI-eligible chats handled by the agentic AI system under the supervision of human workers and classify them by escalation type. Each type captures a distinct failure mode of the agentic AI system:

1) *Algorithm-triggered technical escalations* account for 4,879 chats, or 44.1 percent of the sample. These occur when the monitoring algorithm detects a customer inquiry that exceeds the capabilities of the agentic AI system (see Online Appendix B for details of the monitoring algorithm).

2) *Algorithm-triggered emotional escalations* account for 954 chats, or 8.6 percent of the sample. These occur when the algorithm detects customer frustration or doubts about the service encounter (see Online Appendix B for details of the monitoring algorithm).

3) *Human-initiated escalations* account for 1,362 chats, or 12.3 percent of the sample. These occur when the human supervisor takes over before algorithm-triggered escalation, often in response to early signs of customer frustration or emerging technical difficulty.

4) The remaining 3,874 chats, or 35.0 percent of the sample, involve *no escalation* and are resolved entirely by the agentic AI system.

To construct counterfactuals, we match each AI-eligible chat handled by agentic AI under the supervision of a treated worker to a comparable chat handled entirely by a control worker. We implement one-to-one nearest-neighbor matching using twelve variables that capture chat, customer, and worker characteristics (Rosenbaum 2002). Details of this matching are provided in Appendix C. As Table A5 shows, the matched control chats closely resemble treated chats within each escalation type. We then conduct chat-level regressions separately by subsample:

$$Y_i = \beta_0 + \beta_1 \, Treat_i + \Gamma_1 \boldsymbol{X}_i^{Cust} + \Gamma_2 \boldsymbol{X}_i^{Worker} + \Gamma_3 \boldsymbol{X}_i^{Chat} + \epsilon_i, \qquad (2)$$

Here, $Y_i \in \{\ln(P_d), Q_c, Q_r\}$ denotes the performance outcome for a given chat $i$. We log-transform chat duration due to its relatively high variance. $Treat_i$ equals one if chat $i$ is handled by a treated worker, and zero otherwise. The coefficient $\beta_1$ captures the average treatment effect. $\epsilon_i$ is the idiosyncratic error term. Robust standard errors are clustered at the worker level. The vectors $\boldsymbol{X}_i^{Cust}$, $\boldsymbol{X}_i^{Worker}$, and $\boldsymbol{X}_i^{Chat}$ control for observable customer, worker, and chat characteristics, including customer tenure and VIP status, worker demographics and pretreatment performance, and temporal and contextual factors such as date, hour of day, and issue category.

### 5.2.2 Effectiveness of Human-in-the-Loop Interventions by Escalation Type

Table 5 reports the results of the subsample analyses. We begin with algorithm-triggered technical escalations in Panel A. These escalations arise when customer inquiries move beyond agentic AI's capabilities during chat sessions. Relative to comparable chats handled entirely by human workers in the control condition, chat duration increases by 19.1 percent ($p < 0.001$). Service quality, however, is preserved: the estimated effects on retrial rates and customer ratings are statistically indistinguishable from those of fully human-handled chats. On average, human intervention can therefore compensate for the technical capability limitations of agentic AI, albeit at the cost of additional time.

The pattern is markedly different in Table 5 Panel B, which reports the results for algorithm-triggered emotional escalations. These escalations arise when the monitoring algorithm detects customer frustration or skepticism about the service encounter. Relative to comparable chats handled entirely by human workers in the control condition, chat duration increases by 40.8 percent ($p < 0.001$), the retrial rate rises by 6 percentage points ($p < 0.01$), and customer ratings decline by 0.928 points ($p < 0.001$). These results indicate that once emotional deterioration has accumulated, human intervention is much less effective at restoring service quality. These chats are not only longer but also have worse objective and subjective service quality.

Table 5 Panel C reports the results for human-initiated escalations and provides a useful contrast. In these chats, human workers proactively intervene before algorithm-triggered escalation. Relative to chats handled entirely by human workers in the control condition, chat duration increases by 9.5 percent ($p < 0.01$), which is smaller than in either type of algorithm-triggered escalations. The retrial rate declines by 3.2 percentage points ($p < 0.1$), and although customer ratings still fall, the decline of 0.524 points ($p < 0.01$) is substantially smaller than in emotionally escalated chats. These findings suggest that discretionary intervention initiated by humans can effectively resolve customer issues and partially reverse the emotional deterioration arising from unsuccessful agentic AI service encounters.

Table 5 Panel D reports the results for chats with no escalation and thus provides a benchmark for the performance of the agentic AI system with no human intervention. These chats exhibit the largest gain in service speed, with chat duration declining by 64.6 percent ($p < 0.001$). However, they also experience a substantial drop in customer ratings of 0.858 points ($p < 0.001$), which lies between the larger decline for emotional escalations and the smaller decline for human-initiated escalations. The retrial rate, by contrast, is statistically indistinguishable from that of chats handled entirely by human workers in the control condition. These results imply that the decline in customer ratings likely reflects differences in the communication styles of the agentic AI system and human workers rather than resolution effectiveness.

### 5.2.3 Why Emotional Escalations Are Hard to Recover: Reduced Human Intervention Effort

To understand why human-in-the-loop intervention is less effective in emotionally escalated chats, we examine workers' intervention effort after escalation. Using session logs, we construct four process measures: *average human response delay, number of human chat rounds, share of human rounds in total chat rounds, and number of human messages*. We also use an LLM-based evaluation framework to assess workers' chat content along four dimensions: *information seeking*, *solution provision*, *empathy*, and *proactivity*.[3] All process and content measures are constructed using only post-escalation chat rounds and therefore capture chat contents contributed exclusively by human workers after their takeover of escalated chats.

[3] The detailed technical setup, prompt engineering, and human validation procedures of LLM-based evaluations are reported in Online Appendix D.

Table 6 shows a clear divergence in intervention effort. Relative to other escalation types, workers intervening in chats with algorithm-triggered emotional escalations respond more slowly and participate less, as indicated by fewer messages, fewer human chat rounds, and a smaller share of total chat rounds. Because the longer response delays are not accompanied by greater textual output, this pattern is more consistent with worker disengagement than with careful deliberation, suggesting a weaker willingness to invest in recovery.

The content measures point to the same conclusion. Across the four dimensions, workers intervening in chats with algorithm-triggered emotional escalations show lower initiative in information seeking, solution provision, and proactivity. Empathy is the only dimension that remains similar across the three escalation types. These findings suggest that when workers are prompted by the algorithm to intervene in emotionally escalated chats, they devote much less effort than in other types of escalated chats.

This pattern is consistent with a mechanism of effort reduction under low perceived recoverability, in line with the theory of *learned helplessness* (Maier and Seligman 1976). When the algorithm fails to detect early signs of customer frustration in a chat, the interaction may continue through unfruitful exchanges without reaching a resolution. By the time the algorithm triggers human intervention, negative sentiment may already be entrenched, leaving little room for recovery. Under these conditions, workers may reduce effort, and emotionally escalated chats therefore become longer without corresponding improvements in resolution, which helps explain the higher retrial rates and lower customer ratings reported in Table 5 Panel B.

#### 5.2.4 When Human-in-the-Loop Interventions Are Effective: Timing Matters

The preceding analysis shows that human-in-the-loop interventions in AI-eligible chats are not uniformly effective. Rather, their effectiveness depends on the nature of the AI service failure and whether intervention is triggered by the algorithm or proactively initiated by human supervisors. The central operational question, therefore, is not simply whether firms should keep humans in the loop, but how human labor should be organized around that loop.

One implication is that earlier human intervention appears particularly valuable in emotionally charged customer interactions. Figures A5 and A6 illustrate this pattern descriptively: algorithm-triggered emotional escalations tend to occur in later stages of the chat, when substantial negative sentiment has already accumulated, whereas both algorithm-triggered technical escalations and human-initiated escalations occur much earlier, when negative customer sentiment is still moderate.

Table 5 shows that these timing differences are associated with substantial variation in service quality outcomes. Chats with algorithm-triggered emotional escalations (Panel B) experience larger declines in both objective and subjective service quality measures. By contrast, chats with human-initiated escalations (Panel C) exhibit improvements in objective service quality, as reflected in lower retrial rates, although customer ratings still decline moderately. Similarly, chats with algorithm-triggered technical escalations (Panel A) are able to achieve service quality levels comparable to those in the control group. Taken together, these comparisons suggest that proactive intervention does not fully offset the decline in customer ratings, but it substantially limits the damage relative to waiting for the algorithm to escalate a chat that has already deteriorated.

Compared to the timing of escalations, whether an escalation is initiated by humans appears to matter less for workers' intervention effort. This can be seen by comparing intervention effort between chats with human-initiated escalations and those with algorithm-triggered technical escalations in Table 6. Specifically, Panel A shows that number of human chat rounds and number of human messages are similar across the two escalation types, with differences insignificant at the 5% level. Likewise, Panel B shows that intervention effort in terms of information seeking, solution provision, and proactivity is also broadly comparable across the two escalation types.

In Alibaba's internal data structure, human-initiated escalations are not explicitly classified as technical or emotional. However, based on our reading of the chat records, these escalations appear to arise from both technical and emotional issues. Taken together, these patterns suggest that the primary advantage of human-initiated intervention lies in its earlier timing: early intervention keeps the interaction in a state where negative customer sentiment remains moderate, leading workers to perceive the chat as more recoverable and therefore worthy of sustained effort.

### 5.3 Managerial Implications on AI Supervision: Specialize or Integrate?

Our findings on the divergent performance of human-in-the-loop interventions across different types of escalations point to an organizational tradeoff between specialization and integration. Under a specialization approach, one group of human workers would primarily monitor AI-eligible chats and intervene when necessary, while another group would focus on resolving AI-ineligible chats from start to finish. The main advantage of this specialization approach is stronger supervisory performance: specialized AI supervisors may detect emotional deterioration earlier, intervene more quickly and willingly, thereby improving the recovery from negative customer experiences. This design is particularly attractive when customer frustrations are frequent or when delayed intervention is costly. Its downside, however, is that workers who no longer handle frontline service tasks may gradually lose contextual knowledge of customer problems and experience skill erosion. In addition, because these workers primarily engage with escalated chats, they are continuously exposed to frustrated customers following unsuccessful AI interactions, which may impose substantial emotional strain and increase the risk of emotional exhaustion.

An integration approach keeps the same workers engaged in both AI supervision and frontline service. Its main advantage is that it generates the positive spillover on AI-ineligible chats, as we have shown in Table 4. To explain this spillover effect, we explore treated workers' multitasking behavior in terms of how they handle concurrent chats. Table 7 shows that when workers in the treated condition handle AI-ineligible chats, they switch away from the focal chat less often, spend less total time away from it, and devote a smaller share of chat time to other chats. Because these workers continue to handle chats from start to finish, they also appear better able to direct their attention toward AI-ineligible cases that require the greatest degree of human judgment. Integration therefore helps preserve broad service expertise while keeping supervisors close to the realities of frontline work. The drawback is that integrated roles may delay proactive human intervention in emotionally sensitive chats, where deterioration becomes increasingly difficult to reverse over time. However, this limitation may be mitigated if the monitoring algorithm can be trained to detect emerging customer emotional deterioration more promptly.

In sum, the appropriate process design for human-in-the-loop interventions depends on how firms balance these competing considerations, as neither a fully specialized approach nor a fully integrated structure dominates across all dimensions.

### 5.4 Robustness Checks

**Potential Endogeneity from Changes in Concurrency and Issue Composition.** One concern is that the deployment of agentic AI may have changed either the concurrency level of workers or the composition of issues assigned to treated workers, such that the estimated treatment effects capture shifts in workload or task difficulty rather than the effect of the technology itself. Several features of the setting help address this concern. The chat-routing algorithm of the platform remained unchanged throughout the study period and assigned incoming chats independently of workers' real-time performance and the number of chats they had previously resolved. In addition, Table A3 shows that the distribution of issue categories is stable across the pretreatment and treatment periods, indicating that the composition of cases did not change systematically after deployment. We further regress worker concurrency on treatment status separately for different chat types. As reported by Table A2, across all specifications, the estimated effects on concurrency are statistically insignificant. Taken together, these findings suggest that the introduction of agentic AI did not systematically affect either the number of simultaneous chats handled by treated relative to control workers or the mix of issues they received. This pattern makes it unlikely that the main results are driven by changes in workload allocation, multitasking intensity, or issue composition.

**Alternative Unit of Analysis.** For our main analysis, our observation unit is at the worker-day level instead of worker-hour level partly because the workers generally do not work a full day. Nevertheless, as a robustness check, we re-estimate the treatment effects at the worker-hour level instead of the worker-day level. This alternative specification uses a finer temporal unit of analysis and therefore allows us to assess whether the main findings are sensitive to the level of aggregation adopted in the baseline models. The results in Table A4 are qualitatively consistent with those from the worker-day analysis. For AI-eligible chats, the same speed–quality trade-off remains: treatment significantly increases service speed but lowers service quality. The only difference is that under this specification, the retrial rate of AI-eligible chats increases significantly, which is consistent with the sign of the main result in Table 3. For AI-ineligible chats, the spillover pattern also persists, with faster service speed and increased service quality.

**Replication Tests of LLM-Based Measures.** To assess the robustness of the final prompts, we conducted a replication test by randomly sampling 500 chats for each use case and repeating the evaluation procedure 10 times with the same LLM and identical hyperparameter settings. Across all use cases, both average accuracy and Cohen's Kappa score exceeded 0.90, indicating highly stable labeling performance (Landis and Koch 1977). We further evaluated model-agnostic robustness by applying an alternative LLM, Qwen3-235B-A22B, to the same samples; inter-model agreement remained high, with Cohen's Kappa above 0.80. These results suggest that the prompts capture the intended operational constructs rather than overfitting to a single model. Appendix D.3 reports the replication design and full results.

**Human Validations of LLM-Based Measures.** To validate the LLM-generated variables, we conducted a formal human validation using the annotation of post-escalation worker chat content characteristics as an example. From a population of 7,541 chats, we drew a random sample of 500 for manual annotation, exceeding the minimum sample size required for a 95 percent confidence level and a 5 percent margin of error. Two trained research assistants, blinded to both the LLM outputs and the experimental treatment assignments, independently annotated the sampled chats using the same coding rubrics and few-shot examples embedded in the LLM prompts. Disagreements were resolved through discussion to produce consensus labels. Comparing these human labels with the LLM outputs, we find high agreement across all outcome variables: accuracy exceeds 0.85 and Cohen's Kappa exceeds 0.8. These results provide strong support for the reliability of our LLM-based measurement approach. Appendix D.4 reports the validation procedure, full results, and an analysis of the small number of disagreement cases.

## 6. Conclusion and Discussion

This study examines a central question in the deployment of agentic AI in customer service operations: when autonomous AI systems interact directly with customers, under what conditions can human-in-the-loop intervention effectively recover service failures? Using a randomized field experiment on Alibaba's Taobao platform, we show that the performance consequences of agentic AI deployment are fundamentally heterogeneous. On the one hand, agentic AI improves service speed and generates substantial efficiency gains, especially in AI-eligible chats. On the other hand, these gains do not uniformly translate into better service outcomes. While the objective service quality measure, retrial rates, remains largely unchanged, the subjective service quality measure, customer ratings, diverges sharply across chat types, declining in AI-eligible chats but improving in AI-ineligible chats.

These findings contribute to a more nuanced understanding of human-AI collaboration in service operations. Existing research on human-in-the-loop systems has largely focused on technical settings in which AI failures are primarily cognitive and human intervention serves as a safeguard against well-defined errors. Our results show that customer service poses a different challenge. In this setting, failures may involve not only technical mismatch, but also emotional deterioration that unfolds during the chat. Human intervention is effective when the failure is primarily technical, and it is also more effective when workers intervene proactively before customer sentiment has fully deteriorated. By contrast, once the interaction has accumulated substantial negative emotions, the scope for recovery narrows considerably.

The evidence further suggests that this weaker performance of human-in-the-loop intervention in emotional escalations is linked to differences in post-escalation human effort. When workers take over emotionally escalated chats, they contribute fewer messages, account for a smaller share of total chat rounds, and show less initiative in information seeking and solution provision. These patterns are consistent with the idea that workers may conserve effort when they perceive a service failure as difficult to recover. More broadly, the findings indicate that the effectiveness of human-in-the-loop intervention depends not only on whether humans remain in the loop, but also on the state of the chat at the moment of intervention.

The study also highlights an important organizational tradeoff. A specialized supervisory structure might improve the timeliness and quality of intervention in AI-eligible chats, particularly when emotional failures are common and costly. At the same time, an integrated structure can generate positive spillovers by allowing workers to reallocate attention toward AI-ineligible chats that require full human judgment. The appropriate organizational design therefore depends on the firm's operational environment and the relative importance of timely intervention versus broader workload spillovers. In this sense, the performance of hybrid human-AI systems is shaped not only by the capabilities of the AI system itself, but also by how firms redesign human labor around it.

Taken together, our findings suggest that the deployment of agentic AI should be understood not simply as a technological substitution of human labor, but as an opportunity for process redesign that answers questions such as how agentic AI should be supervised, when humans should intervene, and how effectively their effort can restore service failures.

## References

**Altman, D., Yom-Tov, G. B., Olivares, M., Ashtar, S., & Rafaeli, A. (2021).** Do customer emotions affect agent speed? An empirical study of emotional load in online customer contact centers. *Manufacturing & Service Operations Management*, *23*(4), 854-875.

**Bai, B., Dai, H., Zhang, D. J., Zhang, F., & Hu, H. (2022).** The impacts of algorithmic work assignment on fairness perceptions and productivity: Evidence from field experiments. *Manufacturing & Service Operations Management*, 24(6), 3060-3078.

**Bai, B., Zeng, Z., Zhang, D., & Xu, Z. (2026).** The impact of generative ai search on content-sharing platforms. *Available at SSRN 6171026*.

**Benjaafar, S., Wang, Z., & Yang, X. (2025).** Human in the loop automation: Ride-hailing with remote (tele-) drivers. *Management Science*, *71*(3), 2527-2543.

**Brynjolfsson, E., Li, D., & Raymond, L. (2025).** Generative AI at work. *The Quarterly Journal of Economics*, *140*(2), 889-942.

**Chen, Z., & Chan, J. (2024).** Large language model in creative work: The role of collaboration modality and user expertise. *Management Science*, *70*(12), 9101-9117.

**Cui, R., Li, M., & Zhang, S. (2022).** AI and procurement. *Manufacturing & Service Operations Management*, 24(2), 691-706.

**Goes, P. B., Ilk, N., Lin, M., & Zhao, J. L. (2018).** When more is less: Field evidence on unintended consequences of multitasking. *Management Science*, *64*(7), 3033-3054.

**Hu, K., Allon, G., & Bassamboo, A. (2022).** Understanding customer retrials in call centers: Preferences for service quality and service speed. *Manufacturing & Service Operations Management*, *24*(2), 1002-1020.

**Landis, J. R., & Koch, G. G. (1977).** An application of hierarchical kappa-type statistics in the assessment of majority agreement among multiple observers. *Biometrics*, 363-374.

**Lebovitz, S., Lifshitz-Assaf, H., & Levina, N. (2022).** To engage or not to engage with AI for critical judgments: How professionals deal with opacity when using AI for medical diagnosis. *Organization Science*, 33(1), 126-148.

**Long, Z., Tezcan, T., & Zhang, J. (2024).** Routing and staffing in customer service chat systems with generally distributed service and patience times. *Manufacturing & Service Operations Management*, *26*(5), 1674-1691.

**Maier, S. F., & Seligman, M. E. (1976).** Learned helplessness: theory and evidence. *Journal of Experimental Psychology: General*, *105*(1), 3.

**Marinova, D., Singh, S. K., & Singh, J. (2018).** Frontline problem-solving effectiveness: A dynamic analysis of verbal and nonverbal cues. *Journal of Marketing Research*, *55*(2), 178-192.

**McKinney, S. M., Sieniek, M., Godbole, V., Godwin, J., Antropova, N., Ashrafian, H., ... & Shetty, S. (2020).** International evaluation of an AI system for breast cancer screening. *Nature*, *577*(7788), 89-94.

**Ni, X., Wang, Y., Feng, T., Lu, L. X., Wang, Y., & Zhou, C. (2026).** Generative AI in Action: Field Experimental Evidence from Alibaba's Customer Service Operations. *arXiv preprint arXiv:2603.29888*.

**Noy, S., & Zhang, W. (2023).** Experimental evidence on the productivity effects of generative artificial intelligence. *Science*, 381(6654), 187-192.

**Ransbotham, S., Kiron, D., Khodabandeh, S., Iyer, S., and Das, A.** The Emerging Agentic Enterprise: How Leaders Must Navigate a New Age of AI, *MIT Sloan Management Review* and Boston Consulting Group, November 2025.

**Rank, J., Carsten, J. M., Unger, J. M., & Spector, P. E. (2007).** Proactive customer service performance: Relationships with individual, task, and leadership variables. *Human Performance*, *20*(4), 363-390.

**Raub, S., & Liao, H. (2012).** Doing the right thing without being told: Joint effects of initiative climate and general self-efficacy on employee proactive customer service performance. *Journal of Applied Psychology*, *97*(3), 651.

**Rosenbaum, P. R. (2002).** Covariance adjustment in randomized experiments and observational studies. *Statistical Science*, *17*(3), 286-327.

**Sheridan, T. B., & Parasuraman, R. (2005).** Human-automation interaction. *Reviews of Human Factors and Ergonomics*, *1*(1), 89-129.

**Sun, J., Zhang, D. J., Hu, H., & Van Mieghem, J. A. (2022).** Predicting human discretion to adjust algorithmic prescription: A large-scale field experiment in warehouse operations. *Management Science*, 68(2), 846-865.

**Wang, L., Huang, N., Hong, Y., Liu, L., Guo, X., & Chen, G. (2023).** Voice-based AI in call center customer service: A natural field experiment. *Production and Operations Management*, *32*(4), 1002-1018.

**Xu, F., Hou, J., Chen, W., & Xie, K. (2025).** Generative AI and organizational structure in the knowledge economy. *arXiv preprint arXiv:2506.00532*.

**Xu, Y., Dai, H., & Yan, W. (2024).** Identity disclosure and anthropomorphism in voice chatbot design: A field experiment. *Management Science*, forthcoming.

**Zhang, S., & Narayandas, D. (2026).** Engaging customers with AI in online chats: Evidence from a randomized field experiment. *Management Science*, *72*(1), 73-95.

## Table 1. Variable Definitions and Summary Statistics

| Variables | Definition | Obs. | Mean | SD |
|---|---|---|---|---|
| **Worker Demongraphics** | | | | |
| treat | 1 if the AI system was deployed for the worker, and 0 otherwise | 647 | 0.53 | 0.50 |
| age | worker age (years) | 647 | 34.12 | 6.86 |
| gender | 1 if male; 0 if female | 647 | 0.19 | 0.39 |
| tenure | years since the worker's initial employment | 647 | 1.04 | 0.67 |
| **Chat Characteristics** | | | | |
| chat duration | time elapsed between chat initaition and termination (seconds) | 680,676 | 462.36 | 405.20 |
| if retrial | 1 if the customer contacted the platform again for the same issue within 7 day | 680,676 | 0.43 | 0.50 |
| customer rating | customer rating of a chat (scaled 1-5) | 115,243 | 3.57 | 1.77 |
| chat rounds | total number of back-and-forth message exchanges in a chat | 680,676 | 7.24 | 5.03 |
| total message count | total number of messages send to the customer in a chat | 680,676 | 13.82 | 8.34 |
| total word count | total number of words sent to the customer in a chat | 680,676 | 202.35 | 157.80 |
| avg response delay | average response delays to customer messages (seconds) | 680,676 | 14.76 | 8.23 |
| total response delay | cumulative response delays to customer messages (seconds) | 680,676 | 107.19 | 98.51 |
| concurrency | number of parallel chats a worker handles or supervises when starting a chat | 680,676 | 3.14 | 1.16 |
| total active time | total active time a worker spends on a chat (seconds) | 680,676 | 248.07 | 240.46 |
| total away time | total time a worker spends away from a chat (seconds) | 680,676 | 212.83 | 250.78 |

## Table 2. Total Effects on All Chats

| | **Total Effects on All Chats** | | |
|---|---|---|---|
| | Service Speed | Service Quality | |
| | ln(Chat Duration) | Retrial Rate | Customer Rating |
| | (1) | (2) | (3) |
| Treat | -0.032**** | 0.003 | 0.055 |
| | (0.007) | (0.004) | (0.033) |
| Time-Variant Controls | Y | Y | Y |
| Worker Fixed Effects | Y | Y | Y |
| Calendar-Day Fixed Effects | Y | Y | Y |
| Observations | 15,172 | 15,172 | 14,467 |
| R-Squared | 0.436 | 0.099 | 0.218 |

*Notes.* * p<0.1; **p<0.05; ***p<0.01; ****p<0.001. Robust standard errors are clustered by workers.

## Table 3. Direct Effects on AI-Eligible Chats

| | **Direct Effects on AI-Eligible Chats** | | |
|---|---|---|---|
| | Service Speed | Service Quality | |
| | ln(Chat Duration) | Retrial Rate | Customer Rating |
| | (1) | (2) | (3) |
| Treat | -0.168**** | 0.023 | -0.412**** |
| | (0.032) | (0.015) | (0.114) |
| Time-Variant Controls | Y | Y | Y |
| Worker Fixed Effects | Y | Y | Y |
| Calendar-Day Fixed Effects | Y | Y | Y |
| Observations | 8,077 | 8,077 | 3,865 |
| R-Squared | 0.131 | 0.097 | 0.211 |

*Notes.* * p<0.1; **p<0.05; ***p<0.01; ****p<0.001. Robust standard errors are clustered by workers.

## Table 4. Spillover Effects on AI-Ineligible Chats

| | Spillover Effects on AI-Ineligible Chats | | |
|---|---|---|---|
| | Service Speed | Service Quality | |
| | ln(Chat Duration) | Retrial Rate | Customer Rating |
| | (1) | (2) | (3) |
| Treat | -0.018** | 0.002 | 0.091*** |
| | (0.007) | (0.004) | (0.033) |
| Time-Variant Controls | Y | Y | Y |
| Worker Fixed Effects | Y | Y | Y |
| Calendar-Day Fixed Effects | Y | Y | Y |
| Observations | 15,159 | 15,159 | 14,319 |
| R-Squared | 0.417 | 0.099 | 0.187 |

*Notes.* * p<0.1; **p<0.05; ***p<0.01; ****p<0.001. Robust standard errors are clustered by workers.

## Table 5. AI-Eligible Chats: Subsample Analysis by Escalation Type

| | Panel A. Algorithm-Triggered Technical Escalation | | |
|---|---|---|---|
| | Service Speed | Service Quality | |
| | ln(Chat Duration) | Retrial Rate | Customer Rating |
| | (1) | (2) | (3) |
| Treat | 0.191**** | 0.004 | -0.007 |
| | (0.017) | (0.009) | (0.089) |
| Customer Characteristics | Y | Y | Y |
| Worker Characteristics | Y | Y | Y |
| Chat Characteristics | Y | Y | Y |
| Observations | 9,758 | 9,758 | 1,667 |
| R-Squared | 0.087 | 0.036 | 0.144 |
| | **Panel B. Algorithm-Triggered Emotional Escalation** | | |
| | Service Speed | Service Quality | |
| | ln(Chat Duration) | Retrial Rate | Customer Rating |
| | (1) | (2) | (3) |
| Treat | 0.408**** | 0.060*** | -0.928**** |
| | (0.035) | (0.021) | (0.176) |
| Customer Characteristics | Y | Y | Y |
| Worker Characteristics | Y | Y | Y |
| Chat Characteristics | Y | Y | Y |
| Observations | 1,908 | 1,908 | 446 |
| R-Squared | 0.170 | 0.049 | 0.177 |
| | **Panel C. Human-Initiated Escalation** | | |
| | Service Speed | Service Quality | |
| | ln(Chat Duration) | Retrial Rate | Customer Rating |
| | (1) | (2) | (3) |
| Treat | 0.095*** | -0.032* | -0.524*** |
| | (0.033) | (0.019) | (0.185) |
| Customer Characteristics | Y | Y | Y |
| Worker Characteristics | Y | Y | Y |
| Chat Characteristics | Y | Y | Y |
| Observations | 2,724 | 2,724 | 418 |
| R-Squared | 0.094 | 0.069 | 0.175 |

| | Panel D. No Escalation | | |
|---|---|---|---|
| | Service Speed | Service Quality | |
| | ln(Chat Duration) | Retrial Rate | Customer Rating |
| | (1) | (2) | (3) |
| Treat | -0.646**** | -0.006 | -0.858**** |
| | (0.021) | (0.009) | (0.093) |
| Customer Characteristics | Y | Y | Y |
| Worker Characteristics | Y | Y | Y |
| Chat Characteristics | Y | Y | Y |
| Observations | 7,748 | 7,748 | 1,208 |
| R-Squared | 0.156 | 0.033 | 0.153 |

*Notes.* *p<0.1; **p<0.05; ***p<0.01; ****p<0.001. Robust standard errors are clustered by workers. Treated AI-eligible chats are first classified by escalation type, and then matched to comparable control chats on customer, worker, and chat characteristics. Each panel reports results for the matched sample within the corresponding escalation type.

## Table 6. (Mechanism) AI-Eligible Chats: Post-Escalation Intervention Effort

| | | Panel A. AI Eligible Chats: Post-Escalation Human Chat Process Characteristics | | | |
|---|---|---|---|---|---|
| | N | Average Human Response Delay | Human Chat Rounds | Human Chat Rounds / Total Chat Rounds | Human Messages |
| | | (1) | (2) | (3) | (4) |
| Human-Initiated Escalation | 1362 | 18.663 | 5.421 | 0.601 | 9.907 |
| | | [17.812–19.514] | [5.153–5.690] | [0.587–0.616] | [9.487–10.328] |
| Algorithm-Triggered Technical Escalation | 4879 | 20.006 | 5.452 | 0.654 | 9.966 |
| | | [19.589–20.423] | [5.318–5.587] | [0.647–0.662] | [9.750–10.183] |
| Algorithm-Triggered Emotional Escalation | 954 | 20.965 | 5.135 | 0.433 | 8.907 |
| | | [19.884–22.045] | [4.831–5.440] | [0.417–0.449] | [8.450–9.364] |
| | | **Panel B. AI Eligible Chats: Post-Escalation Human Chat Content Characteristics** | | | |
| | N | Empathy | Information Seeking | Solution Provision | Proactivity |
| | | (1) | (2) | (3) | (4) |
| Human-Initiated Escalation | 1362 | 1.692 | 1.602 | 1.857 | 0.566 |
| | | [1.664–1.719] | [1.567–1.636] | [1.825–1.890] | [0.538–0.593] |
| Algorithm-Triggered Technical Escalation | 4879 | 1.734 | 1.567 | 1.863 | 0.575 |
| | | [1.719–1.748] | [1.549–1.584] | [1.846–1.881] | [0.561–0.589] |
| Algorithm-Triggered Emotional Escalation | 954 | 1.709 | 1.474 | 1.713 | 0.444 |
| | | [1.675–1.743] | [1.434–1.513] | [1.672–1.754] | [0.410–0.477] |

*Notes.* Reported values are sample means; values in brackets denote 95% confidence intervals. Panel A summarizes human workers' chat process characteristics following intervention in AI-eligible chats, by escalation type. Panel B reports LLM-based measures of human workers' chat content following intervention in AI-eligible chats, by escalation type.

## Table 7. (Mechanism) AI-Ineligible Chats: Worker Multitasking Behavior

| | **AI-Ineligible Chats: Worker Multitasking Behavior** | | |
|---|---|---|---|
| | Number of Chat Switch-Aways | Total Away Time from Focal Chat | Share of Away Time from Focal Chat |
| | (1) | (2) | (3) |
| Treat | -0.150** | -5.995** | -0.008** |
| | (0.060) | (2.502) | (0.003) |
| Time-Variant Controls | Y | Y | Y |
| Worker Fixed Effects | Y | Y | Y |
| Calendar-Day Fixed Effects | Y | Y | Y |
| Observations | 15,159 | 15,159 | 15,159 |
| R-Squared | 0.675 | 0.513 | 0.742 |

*Notes.* * p<0.1; **p<0.05; ***p<0.01; ****p<0.001. Robust standard errors are clustered by workers.

**Agentic AI and Human-in-the-Loop Interventions: Field Experimental Evidence from Alibaba's Customer Service Operations**

**Online Appendix**

## Appendix A. Figures

**Figure A1. Distribution of Worker Tenure at the Time of Experiment**

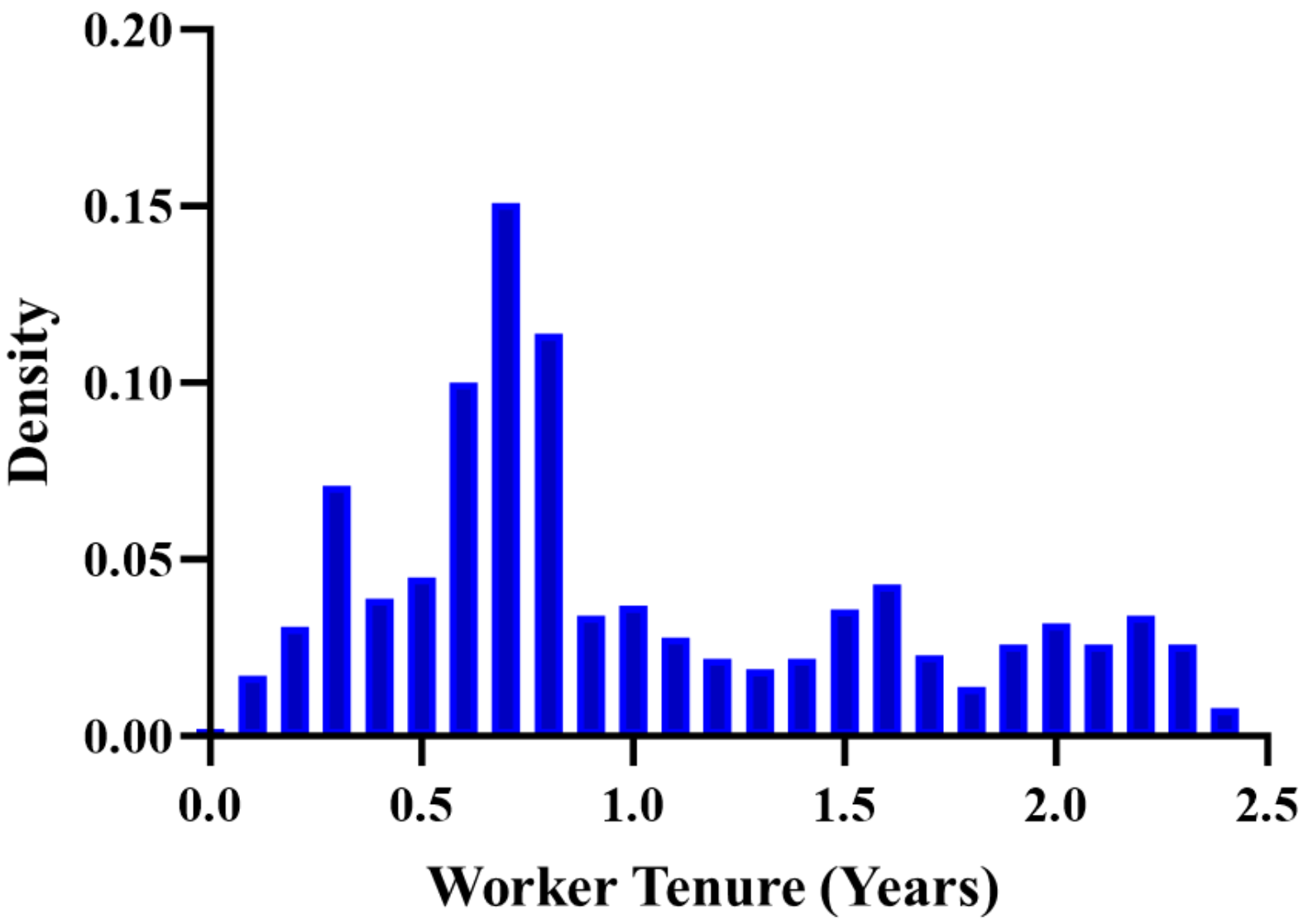

**Figure A2: Service Performance Measures for a Hypothetical Chat Session**

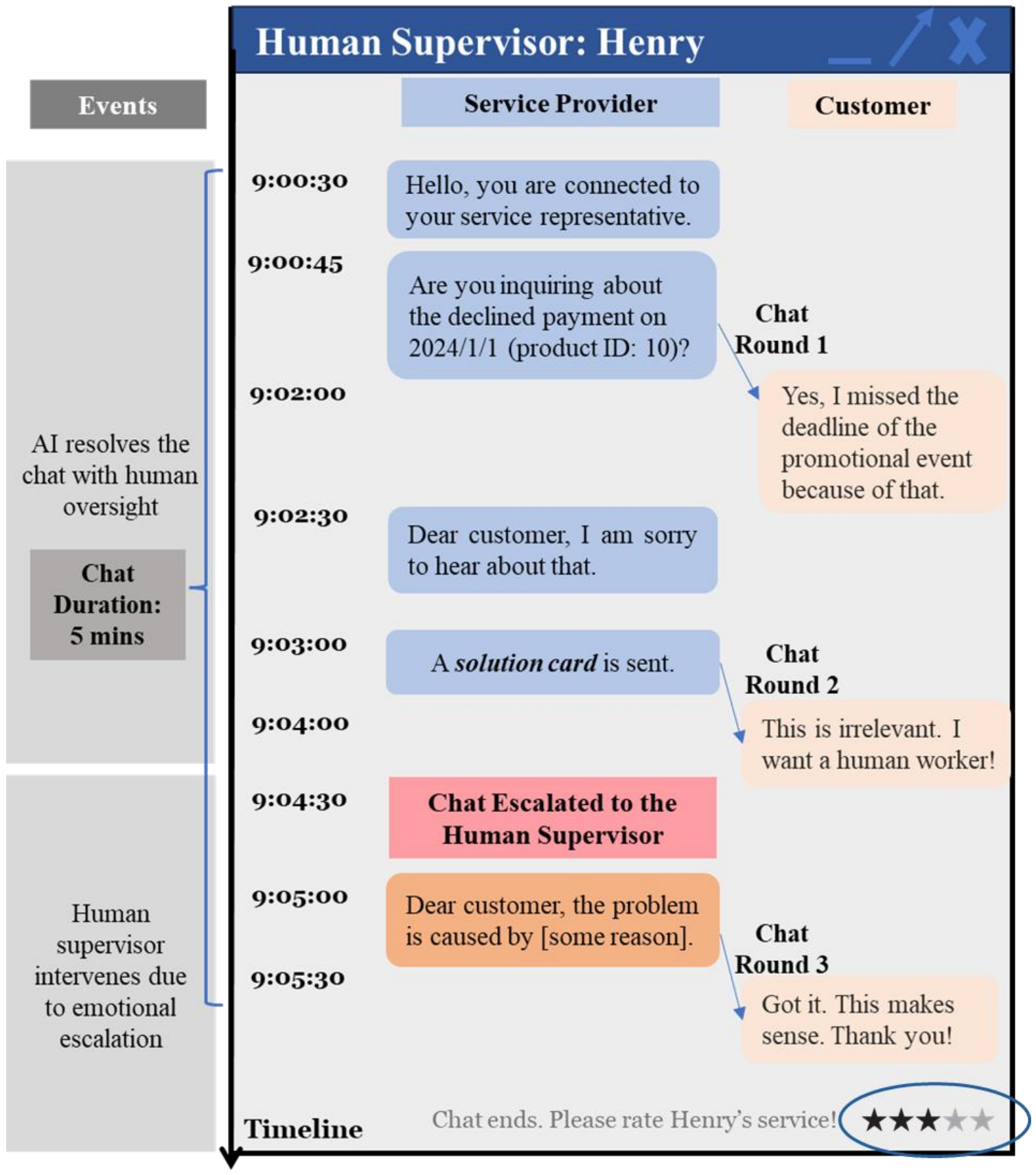

*Notes.* Figure A2 illustrates service performance measures for a hypothetical chat session. The chat started at 9:00:30 with the message "Hello, you are connected to your service representative." The chat ended at 9:05:30 with the message "Got it. This makes sense. Thank you!" In this example, the *chat duration* is 5 minutes. *Customer rating* is 3 out of 5. Chat rounds equal 3, and the total message count is 8.

**Figure A3: Average Daily Rating Response Rates for Treated vs. Control Workers**

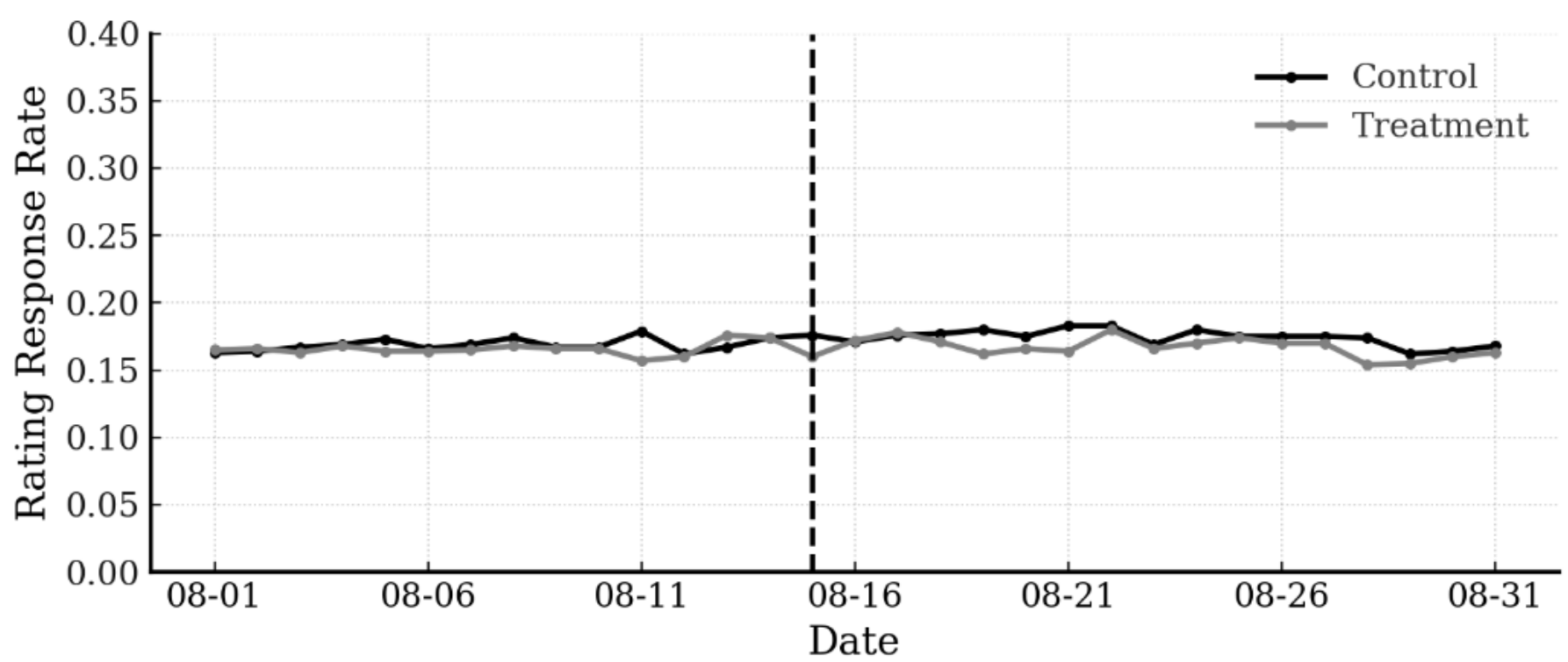

**Figure A4: Pretreatment Distribution of Ratings for Treated vs. Control Workers**

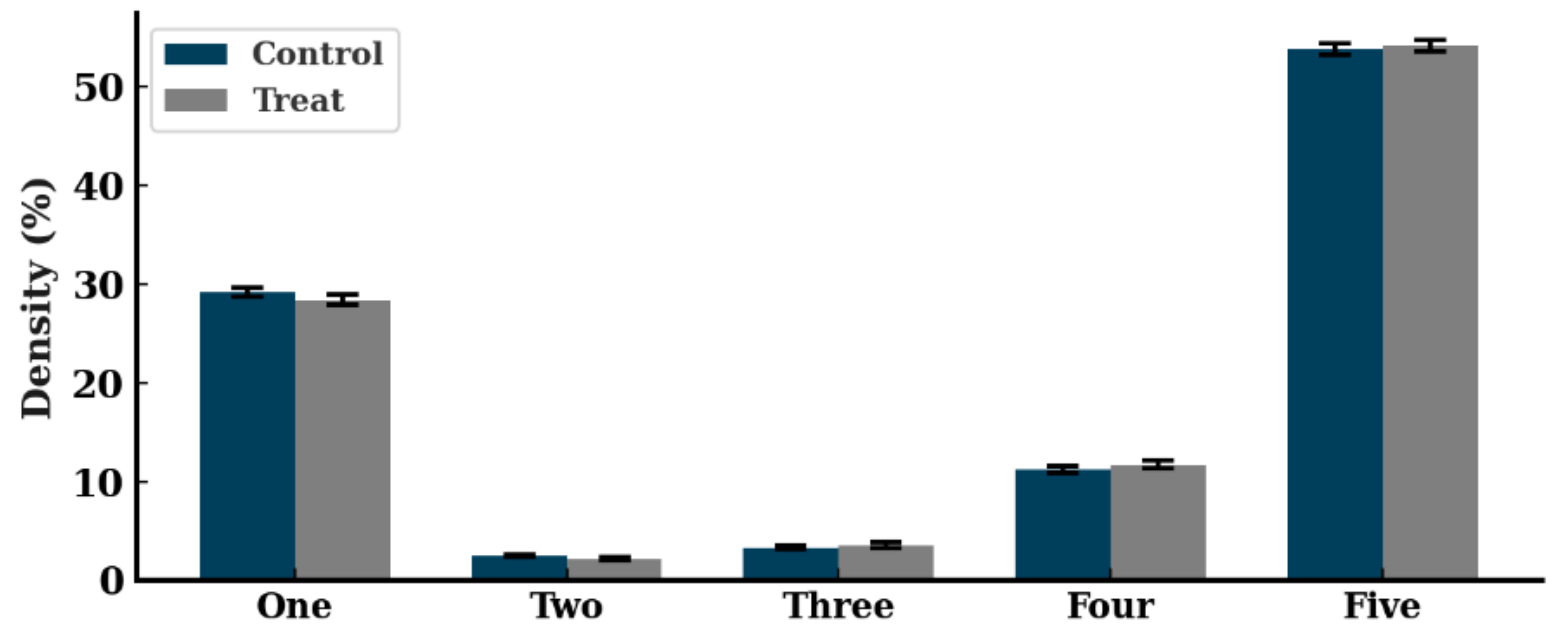

**Figure A5: Density of Chat Escalations Over Chat Progress**

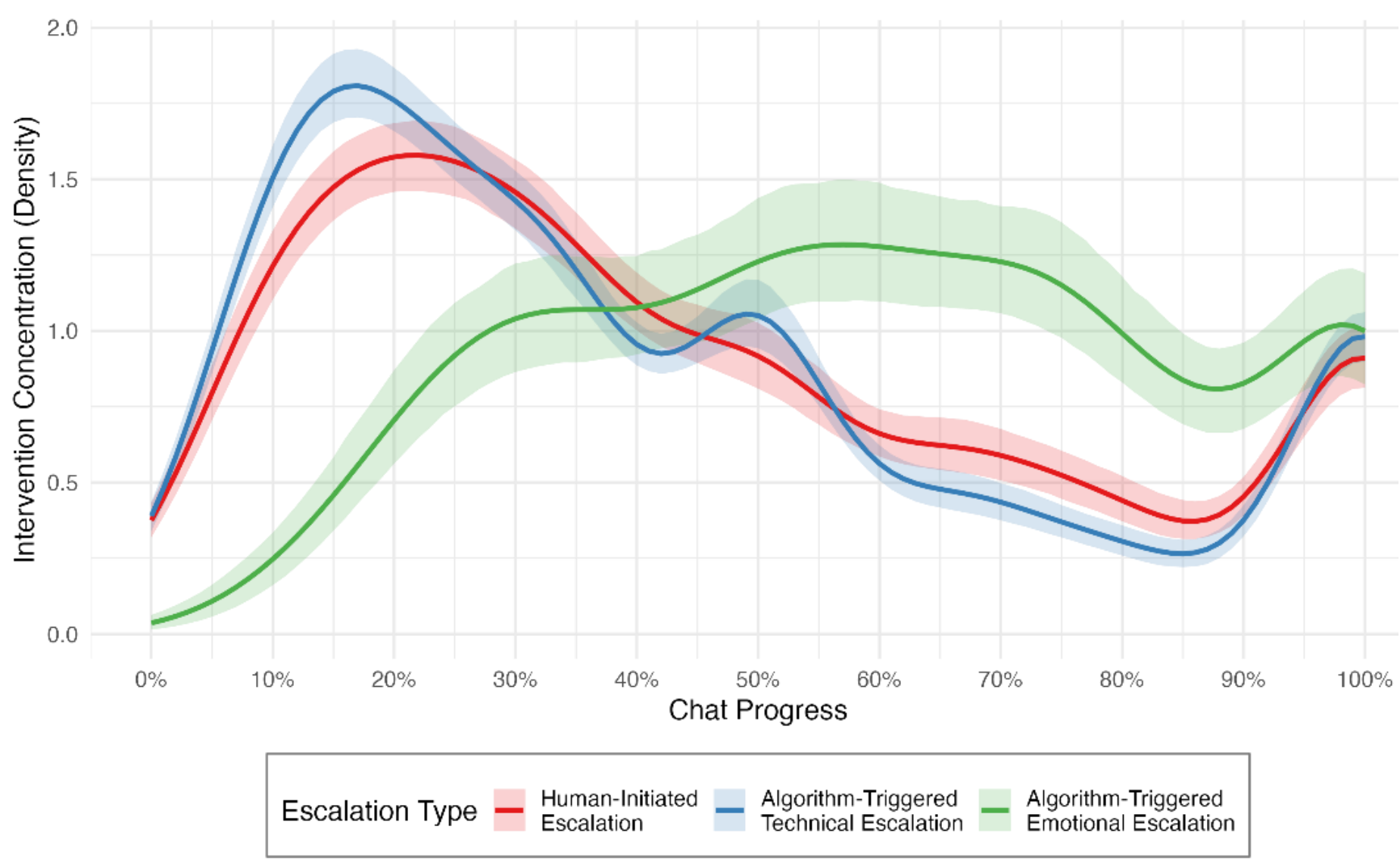

*Notes.* Intervention timing was normalized to [0,1] and summarized using kernel density estimates separately by escalation type. Solid lines show mean bootstrap densities based on 1,000 resamples; shaded areas indicate 95% bootstrap confidence intervals. The y-axis indicates intervention concentration (density).

**Figure A6: Customer Negative Sentiment Over Chat Progress by Escalation Type**

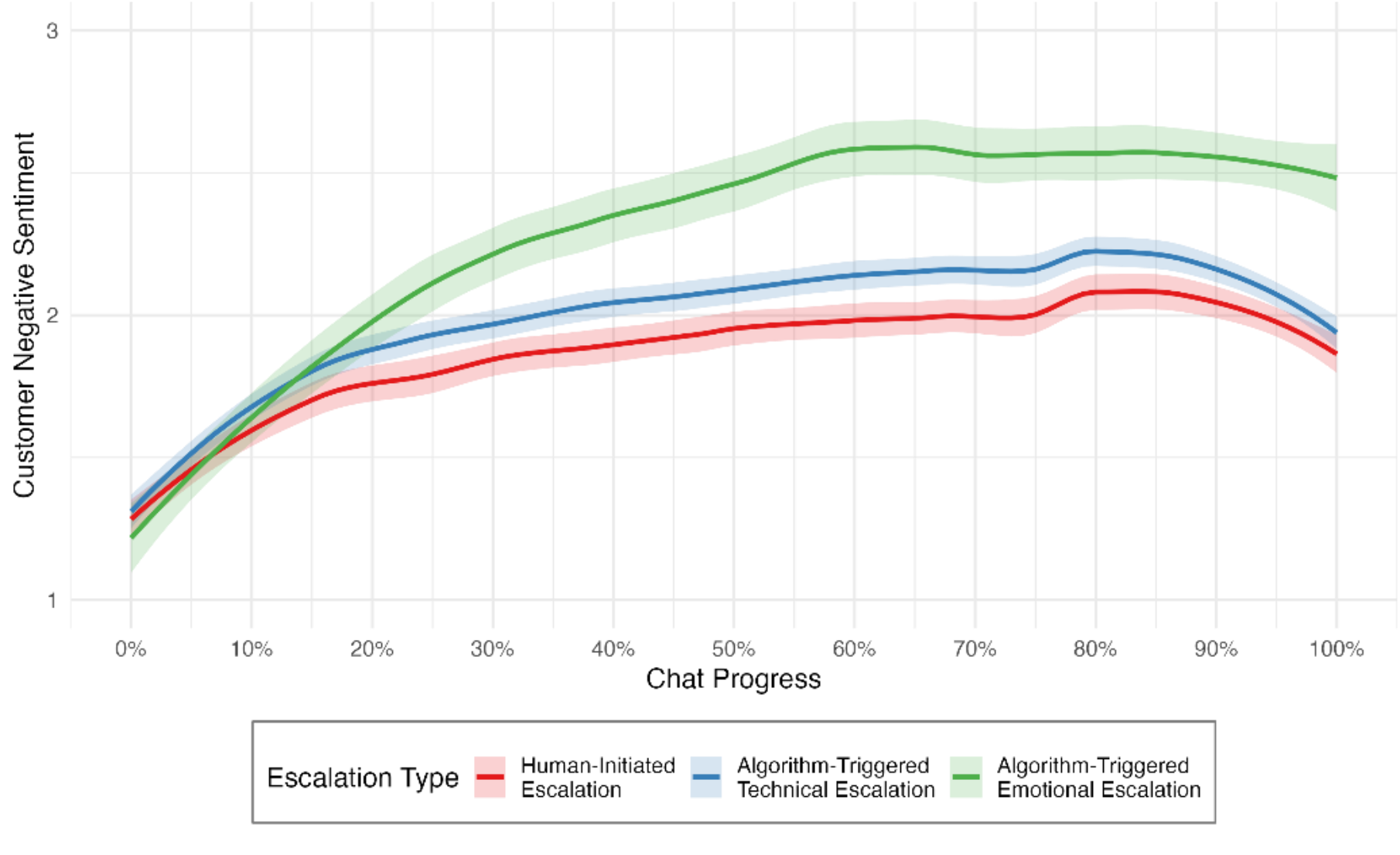

*Notes.* The sentiment trajectories were estimated using LOESS regression separately for each escalation type. Shaded bands indicate 95% confidence intervals. The x-axis shows chat progress normalized to [0,1], and the y-axis shows the customer negative sentiment scores.

# Appendix A. Tables

## Table A1. Balance Check

| | Worker Demographics | | |
|---|---|---|---|
| | treatment | control | p-value |
| N | 302 | 345 | |
| Age | 34.35 | 33.92 | 0.431 |
| | (6.89) | (6.85) | |
| Gender | 0.18 | 0.19 | 0.684 |
| | (0.38) | (0.39) | |
| Tenure (year) | 0.99 | 1.02 | 0.525 |
| | (0.62) | (0.63) | |

| | Chat Characteristics | | |
|---|---|---|---|
| | treatment | control | p-value |
| N | 3320 | 3994 | |
| Chat Duration | 481.51 | 479.31 | 0.420 |
| | (119.16) | (113.94) | |
| Customer Rating | 3.54 | 3.57 | 0.173 |
| | (0.91) | (0.89) | |
| Retrial Rate | 0.44 | 0.45 | 0.152 |
| | (0.04) | (0.05) | |
| Total Chat Rounds | 7.36 | 7.31 | 0.069 |
| | (1.35) | (1.29) | |
| Total Message Count | 13.84 | 13.94 | 0.146 |
| | (2.81) | (3.00) | |
| Average Response Delay | 15.13 | 15.34 | 0.020 |
| | (3.74) | (3.96) | |
| Total Response Delay | 112.03 | 112.43 | 0.602 |
| | (33.45) | (32.26) | |
| Concurrency | 2.84 | 2.82 | 0.236 |
| | (0.75) | (0.76) | |
| Total Active Time | 284.39 | 282.84 | 0.553 |
| | (122.40) | (100.46) | |
| Total Away Time | 200.90 | 200.07 | 0.644 |
| | (76.60) | (77.09) | |
| Rating Response Rate | 0.17 | 0.17 | 0.159 |
| | (0.08) | (0.08) | |

*Notes.* Worker demographics are evaluated at the worker level. Chat characteristics are evaluated at the worker-day level, where each observation corresponds to the set of chats handled by a given worker on a given day during the pretreatment period (August 1–14, 2024).

## Table A2. Effects on Concurrency

| | Concurrency | | |
|---|---|---|---|
| | All Chats | AI-Eligible Chats | AI-Ineligible Chats |
| | (1) | (2) | (3) |
| Treat | -0.005 | 0.004 | -0.003 |
| | (0.018) | (0.031) | (0.019) |
| Time-Variant Controls | Y | Y | Y |
| Worker Fixed Effects | Y | Y | Y |
| Calendar-Day Fixed Effects | Y | Y | Y |
| Observations | 15,172 | 8,077 | 15,159 |
| R-Squared | 0.828 | 0.670 | 0.818 |

*Notes.* * p<0.1; **p<0.05; ***p<0.01; ****p<0.001. Robust standard errors are clustered by workers.

## Table A3. Distribution of Issue Categories: Pretreatment vs. Treatment Periods

| Issue Category | Pretreatment (%) | Treatment (%) |
|---|---|---|
| aftersales | 46.825 | 47.215 |
| shipping | 11.023 | 10.11 |
| risk | 7.846 | 7.603 |
| payment | 6.213 | 7.281 |
| promotion | 6.006 | 5.731 |
| longtail | 5.54 | 5.127 |
| complaint | 5.695 | 6.986 |
| logistics | 5.407 | 4.771 |
| membership | 1.644 | 1.403 |
| account | 1.285 | 1.191 |
| others | 2.517 | 2.582 |

## Table A4. Agent-Hour Level Treatment Effects on Service Performance

| | All Chats | | | AI-Eligible Chats | | | AI-Ineligible Chats | | |
|---|---|---|---|---|---|---|---|---|---|
| | ln(Chat Duration) | Retrial Rate | Customer Rating | ln(Chat Duration) | Retrial Rate | Customer Rating | ln(Chat Duration) | Retrial Rate | Customer Rating |
| | (1) | (2) | (3) | (4) | (5) | (6) | (7) | (8) | (9) |
| Treat | -0.025**** | 0.002 | 0.022 | -0.140**** | 0.025** | -0.420**** | -0.014** | 0.000 | 0.058** |
| | (0.007) | (0.004) | (0.028) | (0.027) | (0.012) | (0.105) | (0.007) | (0.004) | (0.028) |
| Time-Variant Controls | Y | Y | Y | Y | Y | Y | Y | Y | Y |
| Worker Fixed Effects | Y | Y | Y | Y | Y | Y | Y | Y | Y |
| Calendar-Day Fixed Effects | Y | Y | Y | Y | Y | Y | Y | Y | Y |
| Observations | 96,577 | 95,677 | 61,908 | 22,898 | 22,898 | 5,939 | 95,641 | 95,641 | 59,855 |
| R-Squared | 0.112 | 0.023 | 0.059 | 0.058 | 0.036 | 0.137 | 0.110 | 0.023 | 0.047 |

*Notes.* * p<0.1; **p<0.05; ***p<0.01; ****p<0.001. Robust standard errors are clustered by workers.

## Table A5: Balance Check of Matched AI-Eligible Chats by Escalation Type

| | Group | N | Panel A. Customer Characteristics | | | | | |
|---|---|---|---|---|---|---|---|---|
| | | | *Number of Customer Chats* | *Customer Tenure (Years)* | *VIP Indicator* | *Initial Sentiment Score* | *Initial Urgency Score* | *First Message Word Count* |
| No Escalation | Control | 3874 | 2.500 | 7.210 | 0.280 | 1.660 | 1.520 | 8.540 |
| | Treatment | 3874 | 2.420 | 7.120 | 0.270 | 1.660 | 1.520 | 8.310 |
| | p-value | | 0.583 | 0.377 | 0.541 | 1.000 | 1.000 | 0.321 |
| Algorithm-Triggered Technical Escalation | Control | 4879 | 2.880 | 7.940 | 0.280 | 1.840 | 1.650 | 12.950 |
| | Treatment | 4879 | 2.820 | 7.870 | 0.270 | 1.840 | 1.650 | 13.040 |
| | p-value | | 0.690 | 0.405 | 0.330 | 1.000 | 1.000 | 0.742 |
| Algorithm-Triggered Emotional Escalation | Control | 954 | 2.940 | 7.800 | 0.320 | 1.670 | 1.500 | 10.500 |
| | Treatment | 954 | 2.800 | 7.780 | 0.300 | 1.730 | 1.530 | 10.030 |
| | p-value | | 0.686 | 0.890 | 0.399 | 0.101 | 0.184 | 0.540 |
| Human-Initiated Escalation | Control | 1362 | 2.400 | 7.430 | 0.250 | 1.700 | 1.540 | 10.140 |
| | Treatment | 1362 | 2.850 | 7.480 | 0.250 | 1.700 | 1.550 | 10.120 |
| | p-value | | 0.123 | 0.747 | 0.860 | 1.000 | 0.942 | 0.974 |

*Notes*. Number of customer chats are computed from 2024.1 to 2024.7.

| | Group | N | Panel B. Worker Characteristics | | | | |
|---|---|---|---|---|---|---|---|
| | | | *Chat Duration* | *Resolution Rate* | *Customer Rating* | *Concurrency* | *Agent Tenure (Years)* |
| No Escalation | Control | 3874 | 459.780 | 0.610 | 3.330 | 2.880 | 1.050 |
| | Treatment | 3874 | 462.290 | 0.610 | 3.340 | 2.910 | 1.050 |
| | p-value | | 0.132 | 0.599 | 0.190 | 0.105 | 0.927 |
| Algorithm-Triggered Technical Escalation | Control | 4879 | 457.760 | 0.600 | 3.330 | 2.860 | 1.050 |
| | Treatment | 4879 | 459.870 | 0.600 | 3.330 | 2.860 | 1.030 |
| | p-value | | 0.102 | 0.706 | 0.549 | 0.897 | 0.131 |
| Algorithm-Triggered Emotional Escalation | Control | 954 | 463.190 | 0.610 | 3.370 | 2.970 | 1.080 |
| | Treatment | 954 | 464.560 | 0.610 | 3.370 | 2.960 | 1.070 |
| | p-value | | 0.688 | 0.636 | 0.742 | 0.649 | 0.730 |
| Human-Initiated Escalation | Control | 1362 | 465.650 | 0.590 | 3.330 | 2.650 | 1.050 |
| | Treatment | 1362 | 467.150 | 0.590 | 3.300 | 2.650 | 1.030 |
| | p-value | | 0.590 | 0.113 | 0.097 | 0.723 | 0.348 |

| | Group | N | Panel C1. Issue Category (Part 1) | | | | | |
|---|---|---|---|---|---|---|---|---|
| | | | *Aftersales* | *Shipping* | *Risk* | *Payment* | *Promotion* | *Longtail* |
| No Escalation | Control | 3874 | 0.034 | 0.001 | 0.003 | 0.739 | 0.112 | 0.083 |
| | Treatment | 3874 | 0.030 | 0.002 | 0.002 | 0.734 | 0.111 | 0.097 |
| | p-value | | 0.304 | 0.206 | 0.637 | 0.588 | 0.885 | 0.039 |
| Algorithm-Triggered Technical Escalation | Control | 4879 | 0.188 | 0.019 | 0.025 | 0.401 | 0.205 | 0.039 |
| | Treatment | 4879 | 0.195 | 0.020 | 0.021 | 0.409 | 0.206 | 0.035 |
| | p-value | | 0.396 | 0.941 | 0.198 | 0.421 | 0.880 | 0.237 |
| Algorithm-Triggered Emotional Escalation | Control | 954 | 0.103 | 0.011 | 0.025 | 0.566 | 0.116 | 0.025 |
| | Treatment | 954 | 0.108 | 0.009 | 0.036 | 0.582 | 0.113 | 0.020 |
| | p-value | | 0.709 | 0.818 | 0.182 | 0.487 | 0.829 | 0.441 |
| Human-Initiated Escalation | Control | 1362 | 0.164 | 0.019 | 0.024 | 0.554 | 0.098 | 0.022 |
| | Treatment | 1362 | 0.150 | 0.020 | 0.027 | 0.539 | 0.110 | 0.027 |
| | p-value | | 0.317 | 0.890 | 0.542 | 0.419 | 0.316 | 0.287 |

| | Group | N | Panel C2. Issue Category (Part 2) | | | | |
|---|---|---|---|---|---|---|---|
| | | | *Complaint* | *Logistics* | *Membership* | *Account* | *Others* |
| No Escalation | Control | 3874 | 0.004 | 0.002 | 0.000 | 0.008 | 0.013 |
| | Treatment | 3874 | 0.003 | 0.002 | 0.000 | 0.008 | 0.011 |
| | p-value | | 0.352 | 0.808 | 1.000 | 0.795 | 0.465 |
| Algorithm-Triggered Technical Escalation | Control | 4879 | 0.036 | 0.010 | 0.012 | 0.024 | 0.041 |
| | Treatment | 4879 | 0.036 | 0.011 | 0.008 | 0.024 | 0.037 |
| | p-value | | 0.870 | 0.762 | 0.086 | 0.894 | 0.229 |
| Algorithm-Triggered Emotional Escalation | Control | 954 | 0.046 | 0.007 | 0.022 | 0.022 | 0.057 |
| | Treatment | 954 | 0.045 | 0.004 | 0.016 | 0.024 | 0.043 |
| | p-value | | 0.913 | 0.364 | 0.313 | 0.760 | 0.171 |
| Human-Initiated Escalation | Control | 1362 | 0.032 | 0.004 | 0.009 | 0.029 | 0.046 |
| | Treatment | 1362 | 0.033 | 0.005 | 0.009 | 0.034 | 0.046 |
| | p-value | | 0.914 | 0.563 | 1.000 | 0.440 | 0.927 |

## Appendix B. A Description of the Agentic AI System's Architecture

Appendix B provides a detailed description of the AI system deployed in the field experiment. Figure B1 presents the overall architecture of the system. The system relies on a large language model (LLM) as its core computational engine to coordinate multiple generative AI functionalities that autonomously manage online service chats.

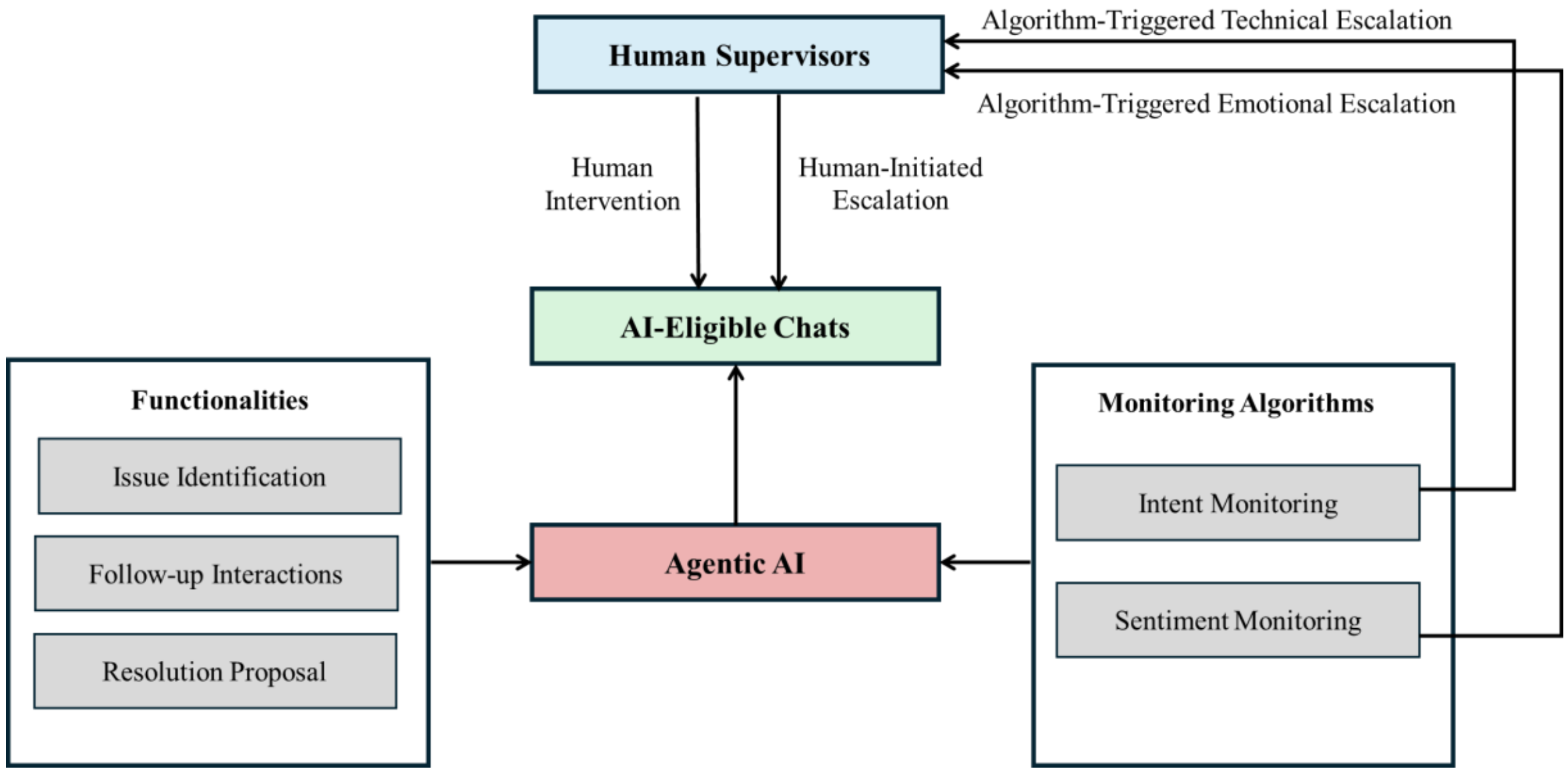

**Figure B1. Architecture of the Agentic AI System**

The architecture of the agentic AI system comprises three primary functionalities. The *Issue Identification* functionality analyzes order information and prior chat history to infer the type of customer issue. The *Follow-up Interactions* functionality engages customers in turn-by-turn dialogue to clarify intent and refine the details of the issue. The *Resolution Proposal* functionality addresses customer issues by following standardized operating procedures (SOPs), which specify predefined actions and response templates for routine and structured service requests.

Together, these functionalities enable the agentic AI system to identify customer issues, conduct clarifying interactions, and implement SOP-based solutions without direct human involvement. Despite these capabilities, the system faces two main challenges that may prevent successful service completion. First, customers may become frustrated by unsuccessful AI attempts, raising doubts about the identity or competence of the service provider. Second, customers may change the intent of their inquiry during the chat, creating issues that exceed the capabilities of the agentic AI system.

To address these limitations, the agentic AI system incorporates two monitoring algorithms. The *Sentiment Monitoring* algorithm continuously evaluates linguistic cues associated with customer frustration and other negative emotions. The *Intention Monitoring* algorithm tracks shifts in customer intent and deviations from AI-eligible issue categories. Together, these algorithms are designed to detect signals of potential service failure.

When these monitoring algorithms detect elevated risk, the system initiates an algorithm-triggered escalation by transferring control of the chat to a human worker. At the same time, the worker monitors AI chats through a dedicated interface and may intervene when they anticipate that the AI system is likely to encounter difficulty, even in the absence of an algorithm-triggered alert. Once intervention happens, the human worker assumes full responsibility for completing the chat.

## Appendix C. Matching Procedures for AI-Eligible Chats

To evaluate differences in the effectiveness of human intervention across escalation types, we must address potential selection bias. A direct comparison between escalated AI-eligible chats and fully human-handled AI-eligible chats may be misleading, because workers may choose to take over chats that are unusually difficult to resolve. To mitigate this concern, we use a matching strategy that pairs chats on a rich set of customer-, worker-, and chat-level covariates.

### C.1 Construction of Matching Variables

We considered a broad set of variables to ensure comparability between treated and control chats in terms of worker, customer, and chat characteristics. For worker characteristics, we included pretreatment averages of chat duration, chat rounds, resolution rate, customer rating, and workload. These variables were calculated over the period from August 1 to August 14, 2024. We also included worker tenure as a demographic characteristic. For customer characteristics, we included VIP status, tenure, and total historical chat volume from January to August 2024.

For chat characteristics, we controlled for the primary problem category of each chat. To capture the initial state of the chat, we used an LLM to annotate the customer's first message and constructed two variables. *Initial Sentiment* measures the customer's initial sentiment on a 1-to-5 scale, where 1 indicates a neutral state and 5 indicates severe negative sentiment. *Initial Urgency* measures the urgency of the request on a 1-to-3 scale, where 3 indicates the highest level of urgency. Details of the construction of these LLM-based variables are reported in Appendix D.

### C.2 Propensity Score Matching

We performed the matching separately for four escalation types: *no escalation*, algo*rithm-triggered technical escalation*, *algorithm-triggered emotional escalation*, and *human-initiated escalation*. For each escalation type, we used the MatchIt package in R to estimate propensity scores with a logistic regression model and then implemented one-to-one nearest-neighbor matching without replacement. To improve covariate balance on the most important potential confounders, we imposed calipers based on standardized distances (std.caliper = TRUE). This procedure was designed to ensure that treated and control chats were closely aligned on observed pre-intervention characteristics. The resulting matching design allows us to identify, for each human-supervised AI chat, a counterfactual fully human-handled chat that began with a highly similar level of customer frustration and urgency, involved the same problem category, and was assigned to a worker with a highly similar historical performance profile.

The original sample comprised 27,469 AI-eligible chats, including 11,507 chats in the treatment group and 15,962 chats in the control group. Within the treated sample, 3,966 chats involved no escalation, 5,179 involved algorithm-triggered technical escalation, 972 involved algorithm-triggered emotional escalation, and 1,390 involved human-initiated escalation. After one-to-one matching, we retained a sample of 22,138 chats, evenly split between treatment and control, with 11,069 chats in each group. In the matched treated sample, 3,874 chats involved no escalation, 4,879 involved algorithm-triggered technical escalation, 954 involved algorithm-triggered emotional escalation, and 1,362 involved human-initiated escalation. Table A5 reports balance checks for the matched sample by escalation type and shows no systematic differences across the 14 key variables between treated chats and their matched controls in any subsample.

## Appendix D. LLM-Based Variable Construction Procedures

Traditional NLP methods, including dictionary-based sentiment analysis, often fail to capture contextual nuance and changes in sentiment over the course of multi-round human-AI chats. In this study, we use an LLM to construct a set of content-based variables from chat transcripts. The remainder of this appendix is organized as follows. Section D.1 describes the technical setup. Section D.2 presents the use cases of the LLM-based analysis. Section D.3 details the prompt engineering procedures. Section D.4 reports the results of human validation.

### D.1 Technical Setup

For all LLM-based analyses, we use Qwen3.5-122B-A10B, a model developed by Alibaba and deployed within the Alibaba intranet. The model is based on a 122-billion-parameter mixture-of-experts architecture, with 10 billion activated parameters per token. It provides strong instruction-following and contextual reasoning capabilities while meeting the requirements of internal deployment at Alibaba. To improve reliability and reduce hallucination in automated annotation, we set the temperature parameter to 0.0 and the top-p sampling parameter to 0.001. By minimizing these hyperparameters, which typically govern output diversity and creativity, we induce the model to select the most probable tokens consistently and thereby generate deterministic and highly stable classification labels.

**D.2 Use Cases of the LLM-Based Analysis**

In this study, the LLM-based analysis serves three purposes: annotating post-escalation worker communication characteristics, constructing variables for the matching procedure in the subsample analysis, and tracing customer negative sentiment over the course of chats by escalation type.

***Case #1.*** We use the LLM to evaluate worker effort along four communication dimensions in 7,541 escalated chats: *information seeking*, *solution provision*, *empathy*, and *proactivity*. The key variables derived from this process are defined as follows.

*Empathy* is measured on a three-point scale that captures the extent to which the worker conveys empathy effectively after escalation, where 1 denotes a mechanical response, 2 denotes general or templated empathy, and 3 denotes empathy tailored to the context of the chat.

*Information Seeking* is measured on a three-point scale that captures the extent to which the worker seeks information after escalation, where 1 denotes limited meaningful information seeking with mainly templated responses, and 3 denotes targeted questioning and active verification.

*Solution Provision* is measured on a three-point scale that evaluates the quality of the solutions provided by the worker after escalation, where 1 denotes vague suggestions or no feasible explanation, and 3 denotes specific and feasible solutions or clear explanations when the issue cannot be resolved.

*Proactivity* is a binary variable that captures whether the worker communicates proactively after escalation, where 0 denotes largely reactive interaction and 1 denotes proactive interaction, such as taking the initiative to explain and follow up.

These measures are widely used in the marketing and operations literature to characterize communication content and service behavior (Rank et al. 2007, Raub et al. 2012, Marinova et al. 2018).

***Case #2.*** To capture the initial state of the 27,469 AI-eligible chats used in the subsample matching analyses, we use an LLM to annotate the first customer message in each chat. We construct *Initial Sentiment*, measured on a five-point scale on which 1 indicates a neutral state and 5 indicates severe negative emotion, and *Initial Urgency*, measured on a three-point scale on which 3 indicates the highest level of urgency. Both measures are widely used in the literature on customer service and human-AI interactions (Altman et al. 2021, Brynjolfsson et al. 2025).

| Service Provider | Chat Round | Customer Message | Service Provider Message | Empathy | Information Seeking | Solution Provision | Proactivity | Initial Sentiment | Initial Urgency | Customer Negative Sentiment |
|---|---|---|---|---|---|---|---|---|---|---|
| Agentic AI | *1* | Hello | Dear, are you want to ask the reason for the closed transaction? | | | | | *1* | *1* | *1* |
| | *2* | No, I paid 653, but the refund only arrived at 598. What's the reason for this? | Dear, I'm deeply sorry for the problem you have met. | | | | | | | *2* |
| | *3* | *Screenshot for refund bill* | SOP: Dear, after verification, the refund order you reported has been successfully processed. There is no issue of the refund. | | | | | - | - | *1* |
| Human Worker | *4* | Why I haven't even open the package? | Dear, please wait for a moment. I will check it for you right here. | | | | | - | - | *2* |
| | *5* | Where is my money? | Is this your order number: ***? | | | | | - | - | *2* |
| | *6* | Yes, I paid 653 for this product. | Dear, thanks for waiting. There is also a service fee which you didn't report for refund. I will help you to refund the left money. | *2* | *2* | *3* | *2* | - | - | *1* |
| | *7* | I have just received the refund, thank you! | Dear, could you please check the balance now? | | | | | - | - | *1* |
| | *8* | Really Thanks | It's my pleasure, dear. This is what I should do and thanks for your understanding. | | | | | - | - | *1* |
| | *9* | | It's very lucky to have met such a gentle and polite customer as you! Wish you a good health and lovely day. | | | | | - | - | - |

**Figure D1: An Example of LLM-Based Variable Construction from Chat Transcripts**

***Case #3.*** To examine the evolution of customer sentiment in AI-eligible chats, we use an LLM to analyze customer sentiment in each chat round. The variable of interest is *Customer Negative Sentiment*, measured on a five-point scale, where 1 indicates neutral sentiment and 5 indicates severe negative sentiment.

Figure D1 illustrates how these three use cases apply to a single AI-eligible escalated chat. The example contains nine rounds. The agentic AI handles the first three rounds but fails to resolve the customer refund discrepancy, which leads to escalation to a human worker who handles the remaining rounds and ultimately resolves the issue. The figure shows, first, how the post-escalation responses of the human worker are used to annotate *empathy*, *information seeking*, *solution provision*, and *proactivity*; second, how the initial customer message is used to construct *Initial Sentiment* and *Initial Urgency*; and third, how *customer negative sentiment* is labeled round by round over the full course of the chat.

For the sentiment evolution analysis, we normalize each chat to a common timeline from 0 percent, corresponding to the initial round, to 100 percent, corresponding to the final round. We then use these normalized timelines to estimate average sentiment trajectories by escalation type with LOESS regression. The resulting curves, shown in Figure A6, trace the evolution of customer negative sentiment over the service process, and the shaded bands represent 95 percent confidence intervals.

### D.3 Prompt Engineering Procedures

The goal of the prompt engineering process was to satisfy two objectives: *accuracy* and *robustness*. *Accuracy* refers to the extent to which the annotations generated by the LLM align with the nuanced judgments of human domain experts. *Robustness* refers to the ability of the LLM to produce deterministic and consistent labels across repeated runs, thereby reducing the stochastic variation inherent in generative models.

Meeting both objectives is difficult because real customer service transcripts are often ambiguous, and many qualitative behaviors are inherently hard to classify without clear conceptual boundaries. For example, a naive LLM may mistake superficial politeness, such as "*Sorry for the inconvenience*," for genuine empathy, or treat a generic phrase such as "*Let me check*" as evidence of proactive problem solving. Customer service chats also contain edge cases and noisy data. Some chats end abruptly, some contain fragmented sentences, and some are so short that they provide little evidence of worker effort. In addition, human workers often rely on templated standard operating procedures rather than fully customized responses, which makes genuine effort difficult to infer. LLMs may also hallucinate in low-information settings or when prompts are weak. In such cases, the model may infer intent that is not actually present, resulting in inconsistent evaluations.

To address these challenges, we incorporated Alibaba's internal templates into the prompt engineering process. These templates are used in the production environment for customer service performance evaluation, automated quality assurance, and staff training. Building on these templates, we implemented a three-stage refinement process.

In the first stage, we revised the prompts to include explicit and mutually exclusive rating criteria. Early iterations showed that the model struggled to distinguish between overlapping concepts, such as politeness versus empathy or asking a question versus proactive behavior. Based on an error analysis of these difficult cases, we introduced additional constraints, such as "*standard apologies alone do not constitute high empathy*" and "*simply saying 'I will check' does not constitute information seeking*," to help the model distinguish more clearly among related concepts.

In the second stage, we moved from zero-shot prompting to few-shot prompting. We first tested zero-shot prompts in which the model was asked to evaluate worker behavior or customer status using only high-level definitions. The outputs showed substantial variation. To reduce hallucination and improve calibration, we assembled a diverse set of representative chat snippets and included the corresponding human evaluations for each snippet in the prompt.

In the third stage, we addressed data quality issues and edge cases. For very short chats or heavily templated responses, we introduced a conservative scoring rule. This rule instructed the model to assign lower scores when sufficient evidence was absent.

To validate the robustness of the final prompts, we conducted a replication test. We randomly sampled 500 chats for each use case and repeated the labeling process 10 times using the same LLM and the same hyperparameter settings. We then calculated average accuracy and Cohen's Kappa. Unlike simple percentage agreement, Cohen's Kappa measures inter-rater reliability while accounting for agreement by chance. Under standard interpretation guidelines, a score above 0.80 indicates strong agreement. According to the replication test, all variables achieved Kappa values above 0.90.

To assess generalizability and model-agnostic robustness, we also evaluated the same sample of 500 chats using an alternative LLM, Qwen3-235B-A22B. Inter-model reliability remained high, with Cohen's Kappa above 0.80.

Taken together, these results provide strong quantitative evidence for the validity of the final prompts and suggest that the prompt engineering strategy captures the intended operational constructs rather than overfitting to a single model. An example of the annotation prompt is provided in Online Appendix E.

### D.4 Human Validation Procedures

To assess the accuracy of the LLM-generated variables against human judgment, we conducted a formal human validation exercise. We first calculated the minimum required sample size for each raw measure based on a 95 percent confidence level and a 5 percent margin of error. To ensure adequate statistical power, we conservatively drew a random sample of 500 chats for manual evaluation for each measure.

| Variable | Raw Sample Size | Validation Sample Size | Label Range | Percentage Agreement | Kappa Score |
|---|---|---|---|---|---|
| Empathy | 7,541 | 500 | 1-3 | 94.99% | 0.88 |
| Information Seeking | 7,541 | 500 | 1-3 | 92.18% | 0.86 |
| Solution Provision | 7,541 | 500 | 1-3 | 93.39% | 0.87 |
| Proactivity | 7,541 | 500 | 0-1 | 94.99% | 0.90 |
| Initial Sentiment | 27,469 | 500 | 1-5 | 91.09% | 0.84 |
| Initial Urgency | 27,469 | 500 | 1-3 | 100% | 1.00 |
| Customer Sentiment | 11,507 | 500 | 1-5 | 87.39% | 0.82 |

**Table D1: Human Validation Results for LLM-Based Annotations**

Two independent research assistants, who were trained in data annotation and blinded to both the LLM outputs and the experimental treatment assignments, manually annotated the 500 sampled chats. They used the same coding rubrics and few-shot examples embedded in the LLM prompts. Any disagreements were resolved through discussion to produce a final consensus label for each chat. We then evaluated the alignment between the human labels and the LLM outputs using overall accuracy and Cohen's kappa.

The validation results indicate a high level of agreement between the human annotators and the LLM. Across all outcome variables, accuracy exceeded 85 percent, and Cohen's Kappa for the comparison between human labels and LLM outputs consistently exceeded 0.8. These findings provide strong empirical support for the accuracy and reliability of the LLM-generated variables and increase confidence in the validity of the measures used in the main econometric specifications.

Although disagreement was rare, the remaining cases are informative because they clarify the boundary conditions of using an LLM as an evaluator. Table D2 groups the main sources of disagreement into three categories: absence of context, implicit sentiment, and data integrity and observability. These discrepancies account for only a very small share of the validation sample. A qualitative review of these edge cases suggests that the errors of the LLM mainly reflect strict adherence to prompt instructions, such as avoiding hallucinated context, rather than a systematic failure of reasoning. Accordingly, these minor limitations in parsing implicit context or handling incomplete data logs are more likely to introduce classical measurement noise than systematic bias.

| Type of Error | Example | Reason |
|---|---|---|
| Absence of Context | LLM Input: {Customer: I can use this payment function for my first order, but why can I not use this function for my following orders?}<br>LLM Output: {Initial Sentiment: 2, Initial Urgency: 1}<br>Human Output: {Initial Sentiment: 1, Initial Urgency: 1} | Because LLM was not provided with the full interaction history, it over-interprets mild inquiries as negative sentiment, whereas human annotators, drawing on broader operational experience, recognize it as a neutral, factual question. |
| Implict Sentiment | LLM Input: {Customer: Hello, why can I not use the red envelopes?}<br>LLM Output: {Initial Sentiment: 1, Initial Urgency: 1}<br>Human Output: {Initial Sentiment: 2, Initial Urgency: 1} | The LLM exhibits a literal bias; the initial greeting ("Hello") misleads the model into classifying the sentiment as entirely neutral (Sentiment 1). However, human annotators, drawing on domain experience, recognize that inquiries about failed promotions ("red envelopes") inherently involve service friction and typically induce a mild sense of frustration or failure (Sentiment 2). |
| Data Integrity & Observability | LLM Input: {Round:2, Customer content: {I can't pay my order!}, Agent content: {Dear, what's the prompt of this error?}, Round:3, Customer content: {The prompt says that my account is under risk.}, Agent content: {Please upload the screen shot to me, dear.}, Round: 4, Customer content: {<screen shot>}, Agent content: {Dear, please click the reason to find solutions.}, Round: 5, Customer content: {I cannot solve this question following the solutions, my problem is different with these examples}, Agent content: {Missing},….}<br>LLM Output: {Empathy: 1, Information Seeking: 2, Solution Provision:1, Proactivity: 0}<br>Human Output: {Empathy: 1, Information Seeking: 2, Solution Provision:1, Proactivity: 1} | Information Asymmetry: Incomplete chat logs (e.g., missing screenshots or omitted agent responses) lead to systematic measurement bias. The LLM fails to infer latent agent effort from context, whereas human experts can utilize domain knowledge to recognize that a proactivity score of "1" is warranted despite the missing record. |

**Table D2: Illustrative Cases of Disagreement Between Human Annotators and LLM**

## Appendix. E. Prompt Example for LLM Annotation

```
System Role
You are a data labeling expert specializing in customer service dialogue analysis.

User Message
[Task Description]
Below is the first round of a dialogue session. Multiple messages from the user and the
agent in this round have been aggregated into a single entry per role.
Specifically:
  - Within the same round (`round_seq = 1`), all agent utterances are merged into one
    `content` (if any).
  - Within the same round (`round_seq = 1`), all user utterances are merged into one
    `content` (if any).
Your Task
  - Analyze the aggregated content of the user and the agent in the first round to
    understand the user's overall initial state when entering the session.
  - Output a structured JSON object for the "Overall User State in the First Round."
General Requirements
  - Base your judgment strictly on the provided text of the first round; do not
    speculate on information outside the dialogue.
  - You may refer to the agent's content to understand the context, but the final
    labels must only describe the user (do not evaluate the agent).
  - The scores must be primarily based on the user's `content`. The agent's `content`
    serves only as auxiliary context and should not directly dictate the user labels.
  - This is an "overall labeling for aggregated first-round content," not message-by-
    message labeling or a summary of the entire session.
  - Output only a JSON object. Do not include additional text, explanations, or
    comments.

I. Input Format Description
The input is a JSON array where each element represents the aggregated content for a
specific role in this round:
  - `role`: "user" or "agent".
  - `round_seq`: Round number (fixed at 1 for this task).
  - `content`: The merged text of all messages sent by that role in this round.

II. Core Labeling Principles
  - Overall State: This is a judgment of the "overall user state in the first round,"
    not a judgment of the single highest or lowest emotional peak.
  - Dominant Tone: If the user's aggregated content contains multiple expressions, the
    overall dominant tone should prevail, rather than just focusing on the most intense
    sentence.
  - Conservative Judgment: If the content is short or information is insufficient, be
    conservative. Do not over-infer.
  - Distinction: A user having a problem does not automatically imply high emotion;
    high emotion does not automatically imply high urgency. Judge Emotion and Urgency
    independently.

III. User Attributes to Evaluate
1) Initial Sentiment (Negative Emotion Intensity, 1-5)
Judge the intensity of negative emotion based on the overall tone and wording in the
user's aggregated first-round content.
  - 1: None / Neutral / Polite: Tone is polite, calm, or neutral. No obvious
    dissatisfaction, complaining, irritability, pressure, or aggression. (e.g., "How do
    I do this?", "Hello").
  - 2: Slight Negative: Slight confusion, disappointment, inconvenience, or minor
    complaining about the issue, but remains restrained. (e.g., "Why can I not order?",
    "Payment keeps failing").
  - 3: Moderate Negative: Clear dissatisfaction, annoyance, or complaining. May include
    repeated emphasis, urging, or blaming, but has not reached strong pressure or
    threats. (e.g., "Why is it not ready yet?", "Too much trouble").
  - 4: High Negative: Strong anger, clear pressure, severe blaming, or intent to
    complain or escalate, but no extreme insults. (e.g., "Are you guys capable?", "Stop
    stalling,", "I want to complain").
  - 5: Extreme Negative: Strongly hostile, aggressive, insulting, or includes serious
    threats. (e.g., "Trash,", "Scammers,", "Lawyer's letter,", "I will expose you").
2) Initial Urgency (User Urgency, 1-3)
Judge the urgency of resolving the issue based on the user's expression.
  - 1: Low: Focused on gathering information or general help. No obvious time pressure.
    (e.g., "How do I do this?", "Cannot order").
  - 2: Medium: Hopes for prompt handling or expresses clear expectations for a fix, but
    without "immediately/now" demands or severe time-sensitive consequences. (e.g.,
    "Please handle this as soon as possible").
  - 3: High: Explicitly states "Urgent," "Right now," or "Immediately," or describes
    clear time-sensitive consequences (e.g., "I need to pay now or I'll miss the
    meeting").
```

IV. Handling Insufficient Information

- If the content of the user is empty, assign `Initial Sentiment = 1` and `Initial Urgency = 1`.
- If user `content` is very short and only states a problem (e.g., "Payment failed"): `Initial Sentiment` usually 2, `Initial Urgency` usually 1, unless specific pressure or urgency signals are present.

V. Few-Shot Examples

Example 1

**Input**

```
[{"role": "agent", "round_seq": 1, "content": "I'm here! How can I help?"}, {"role": "user", "round_seq": 1, "content": "Cannot order, payment failed"}]
```

**Output**

```
{"content": "Cannot order, payment failed", "Initial Sentiment": 2, "Initial Urgency": 1}
```

Example 2

**Input**

```
[{"role": "agent", "round_seq": 1, "content": "Hello, how can I help?"}, {"role": "user", "round_seq": 1, "content": "Why can't I use this coupon?"}]
```

**Output**

```
{"content": "Why can't I use this coupon?", "Initial Sentiment": 2, "Initial Urgency": 1}
```

Example 3

**Input**

```
[{"role": "agent", "round_seq": 1, "content": "Hello, how can I help?"}, {"role": "user", "round_seq": 1, "content": "Why is it not ready yet? Too much trouble. Please handle this as soon as possible."}]
```

**Output**

```
{"content": "Why is it not ready yet? Too much trouble. Please handle this as soon as possible.", "Initial Sentiment": 3, "Initial Urgency": 2}
```

Example 4

**Input**

```
[{"role": "agent", "round_seq": 1, "content": "Hello, how can I help?"}, {"role": "user", "round_seq": 1, "content": "Are you guys capable? I need to pay right now, handle it immediately or I will complain."}]
```

**Output**

```
{"content": "Are you guys capable? I need to pay right now, handle it immediately or I will complain.", "Initial Sentiment": 4, "Initial Urgency": 3}
```

Example 5

**Input**

```
[{"role": "agent", "round_seq": 1, "content": "Hello, how can I help?"}, {"role": "user", "round_seq": 1, "content": "Your trash platform is a scam."}]
```

**Output**

```
{"content": "Your trash platform is a scam.", "Initial Sentiment": 5, "Initial Urgency": 1}
```

VI. Output Format

Please output a JSON object with the following fields:

```
{
  "content": "",
  "Initial Sentiment": 1,
  "Initial Urgency": 1
}
```

- Output only the JSON object. Do not include explanations, reasoning, or additional text.
- All fields must contain values.
- The `content` field should be a direct transcription of the user's aggregated `content` from the first round.

[Dialogue to Analyze]

{dialog_json_str}